%% file: main.tex
\documentclass[sigconf]{acmart}

\usepackage{amsmath,amsfonts}
\usepackage{algorithmic}
\usepackage{threeparttable}
\usepackage{graphicx}
\usepackage{textcomp}
\usepackage{xcolor}
\usepackage{comment}
\usepackage{subcaption}
\usepackage{multirow}
\usepackage{stfloats}
\usepackage{comment}
\usepackage{balance}

\usepackage[font=small]{caption}

\newtheorem{prop}{Proposition}

\usepackage[skip=0pt]{caption}
\usepackage{algorithm}
\usepackage{algorithmic}
\newcommand{\INDSTATE}[1][1]{\STATE\hspace{#1\algorithmicindent}}

\begin{document}
\fancyhead{}

\title{Discrete-time Temporal Network Embedding via Implicit Hierarchical Learning in Hyperbolic Space}

% Compact author format
% \author{Menglin Yang, Min Zhou, Marcus Kalander, Zengfeng Huang, Irwin King}
% \affiliation{%
%   \institution{$^1$Department of Computer Sciences and Engineering, The Chinese University of Hong Kong, \\$^2$Noah's Ark Lab, Huawei Technologies, $^3$Fudan University}
%   %\city{Hong Kong}
%   %\country{China}
% }

% \email{{mlyang, king}@cse.cuhk.edu.hk, {zhoumin27, marcus.kalander}@huawei.com, huangzf@fudan.edu.cn}

% Correct the author list at the top of each page
% \renewcommand{\shortauthors}{Menglin Yang, Min Zhou, Marcus Kalander, Zengfeng Huang, Irwin King}

\author{Menglin Yang}
%\authornote{Work mainly done during an internship at Huawei Noah's Ark Lab.}
\affiliation{%
  \institution{The Chinese University of Hong Kong}
  %\city{Hong Kong}
  %\country{China}
}
\email{mlyang@cse.cuhk.edu.hk}

\author{Min Zhou}
\affiliation{%
  \institution{Noah's Ark Lab, Huawei Technologies}
  %\city{Hong Kong}
  %\country{China}
}
\email{zhoumin27@huawei.com}

\author{Marcus Kalander}
\affiliation{%
  \institution{Noah's Ark Lab, Huawei Technologies}
  %\city{Hong Kong}
  %\country{China}
}
\email{marcus.kalander@huawei.com}

\author{Zengfeng Huang}
\affiliation{%
  \institution{Fudan University}
  %\city{Hong Kong}
  %\country{China}
}
\email{huangzf@fudan.edu.cn}

\author{Irwin King}
\affiliation{%
  \institution{The Chinese University of Hong Kong}
  %\city{Hong Kong}
  %\country{China}
}
\email{king@cse.cuhk.edu.hk}

%\input{each_parts/abstract}
%\keywords{Temporal Network; Hyperbolic Geometry; temporal graph; Graph Convolutional Network}

\begin{abstract}
Representation learning over temporal networks has drawn considerable attention in recent years. 
Efforts are mainly focused on modeling structural dependencies and temporal evolving regularities in Euclidean space which, however, underestimates the inherent complex and hierarchical properties in many real-world temporal networks, leading to sub-optimal embeddings. 
To explore these properties of a complex temporal network, we propose a hyperbolic temporal graph network (HTGN) that fully takes advantage of the exponential capacity and hierarchical awareness of hyperbolic geometry. More specially, HTGN maps the temporal graph into hyperbolic space, and incorporates hyperbolic graph neural network and hyperbolic gated recurrent neural network, to capture the evolving behaviors and implicitly preserve hierarchical information simultaneously. 
Furthermore, in the hyperbolic space, we propose two important modules that enable HTGN to successfully model temporal networks: (1) hyperbolic temporal contextual self-attention (HTA) module to attend to historical states and (2) hyperbolic temporal consistency (HTC) module to ensure stability and generalization. 
Experimental results on multiple real-world datasets demonstrate the superiority of HTGN for temporal graph embedding, as it consistently outperforms competing methods by significant margins in various temporal link prediction tasks. Specifically, HTGN achieves AUC improvement up to 9.98$\%$  for link prediction and 11.4$\%$ for new link prediction.
Moreover, the ablation study further validates the representational ability of hyperbolic geometry and the effectiveness of the proposed HTA and HTC modules. Code is publicly available at {\color{blue!50!black}\url{https://github.com/marlin-codes/HTGN}}.
\end{abstract}



\maketitle

\section{Introduction}
Data describing the relationships between nodes of a graph are abundant in real-world applications, ranging from social networks analysis~\cite{liu2019characterizing}, traffic prediction~\cite{zhao2019tgcn}, e-commerce recommendation~\cite{HCA}, and protein structure prediction~\cite{fout2017protein} to disease propagation analysis~\cite{hgcn2019}. In many situations, networks are intrinsically changing or time-evolving with vertices (including their attributes) and edges appearing or disappearing over time. 
Learning representations of dynamic structures is challenging but of high importance since it describes how the network interacts and evolves, which will help to understand and predict the behavior of the system. 

\begin{figure}[t]
\centering
\includegraphics[width=4.2cm]{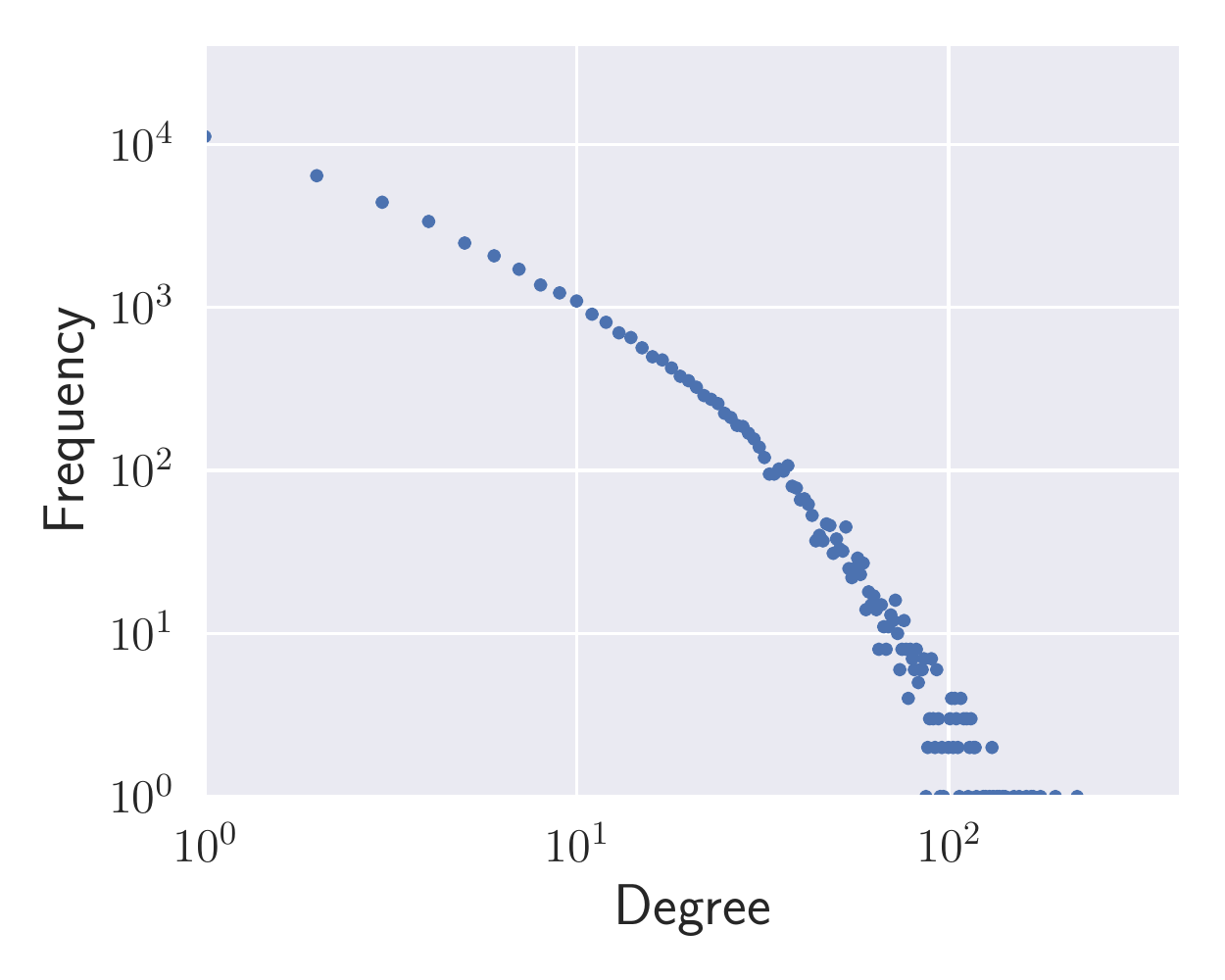}
\includegraphics[width=4.2cm]{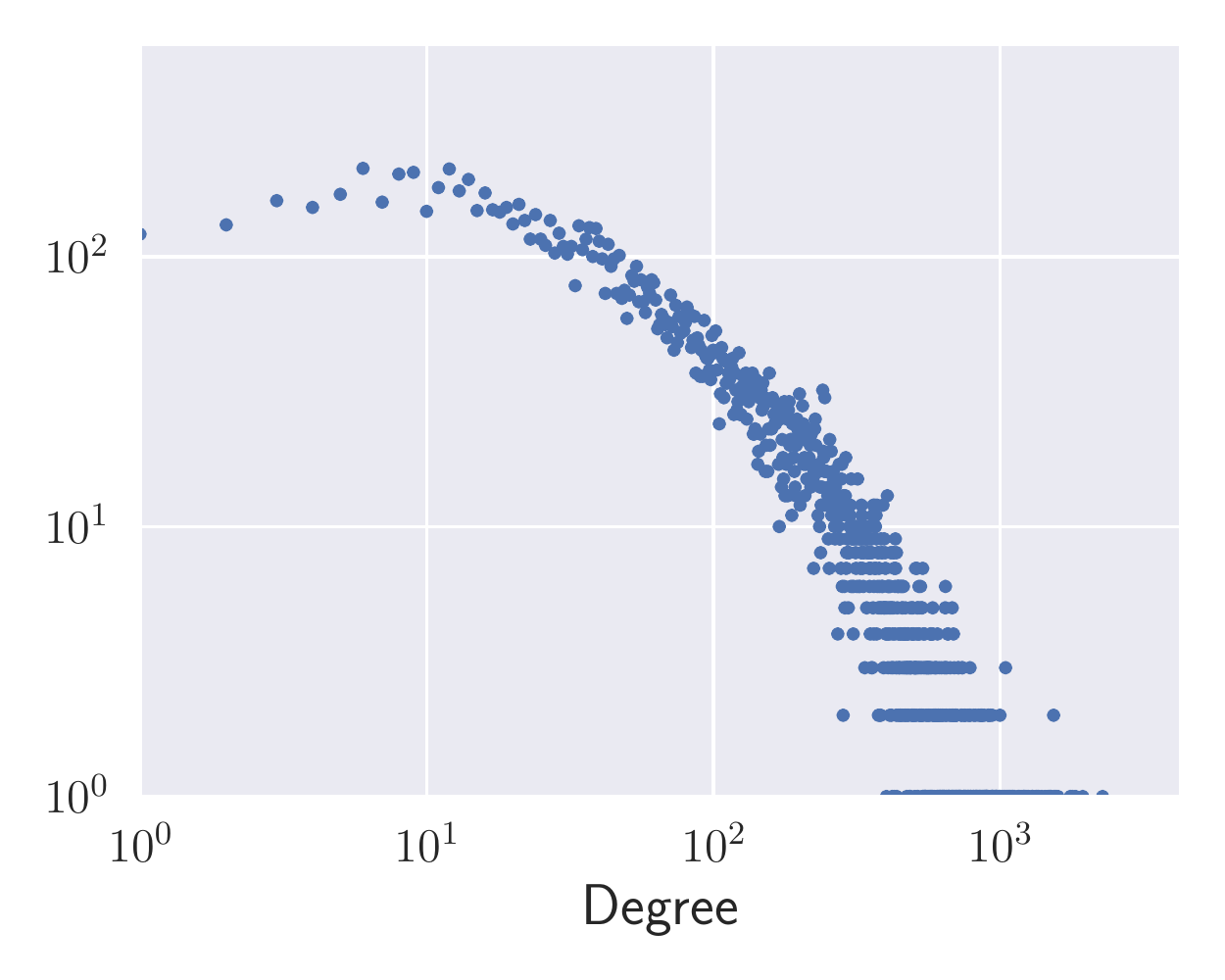}
\caption{Degree distributions of a social network (FB, left) and a citation network (HepPh, right), which are asymptotically power-law distributed.}
\label{fig:degree}
\end{figure}

A number of temporal graph embedding methods have been proposed, which can be divided into two main categories: discrete-time network embeddings and continuous network embeddings~\cite{FeatureNorm2020}. Discrete-time network embeddings are represented in discrete time intervals denoted as multiple snapshots~\cite{sankar2020dysat,pareja2020evolvegcn}.
%DySAT~\cite{sankar2020dysat} proposed temporal attention among multiple timestamps to capture the temporal evolution behaviors; EvolveGCN~\cite{pareja2020evolvegcn} utilized the recurrent neural network (RNN) to preserve temporal patterns in the weights. VGRNN~\cite{hajiramezanali2019VGRNN} extracted a prior distribution based on historical time steps to increase the expressiveness of evolution processes. 
As for continuous-time network embeddings, these can be described as
%typical definition is a 
time-dependent event-based models where the events, denoted by edges, occur over a time span~\cite{tne:ctdn,tne:dyrep,tne:Neural_process}. Essentially, these two schemes both focus on capturing the underlying characteristics of a temporal graph: temporal dependency and topology evolution in Euclidean space. Euclidean space is the natural generalization of
intuition-friendly and visualizable three-dimensional space with appealing vectorial structure, and closed-form formulas of distance and inner-product~\cite{HNN}. However, the quality of the representations
is determined by how well the geometry of the embedding space matches the structure of the data~\cite{gu2019learning}, which triggers one basic question: whether the widely used Euclidean space is the best option for network embedding of an arbitrary temporal graph. 
Several works~\cite{bronstein2017geometric,ying2018hierarchical} show that most of the graph data, e.g., social networks, communication networks, and disease-spreading networks exhibit non-Euclidean latent anatomies that show hierarchical structures and scale-free distributions as illustrated in Figure~\ref{fig:degree}.  %The main reason is that the capacity of the Euclidean space is polynomial, while many real-world networks are complex which present a scale-free distribution and hierarchical patterns. 
This motivates us to rethink (1) whether the Euclidean manifold is the most suitable geometry for graph embedding of this kind of data and (2) is there any more powerful or proper alternative manifold to intrinsically preserve the graph properties, and if it exists, what benefits can it bring to temporal graph embedding? 

Recently, hyperbolic geometry has received increasing attention and achieves state-of-the-art performance in several static graph embedding tasks~\cite{nickel2017poincare,nickel2018learning,hgcn2019,liu2019HGNN,zhang2019hyperbolic}. 
One fundamental property of hyperbolic space is that it expands exponentially and can be regarded as a smooth version of trees, abstracting the hierarchical organization~\cite{2010hyperbolic}. Therefore, the  (approximate) hierarchical structure and tree-like data can naturally be represented by hyperbolic geometry, which instead will lead to severe structural inductive biases and high distortion if directly embedded into Euclidean space. %Furthermore,  % is  when dealing with data with scale-free distribution, the hyperbolic geometry is more natural and causes less distortion. On the other hand, hyperbolic geometry can be a key observation providing a rationale of hierarchical organization. Thus, hyperbolic geometry is able to implicitly model the underlying hierarchical structure of a complex graph.
Despite the recent achievements in hyperbolic graph embedding, attempts on temporal networks are still scant. %, \textcolor{red}{it is not leverage the powerful representation ability of hyperbolic geometry for temporal graph learning algorithms. : (1) One this that}  
To fill this gap, in this work, we propose a novel hyperbolic temporal graph network (HTGN), which fully leverages the implicit hierarchical information to capture the spatial dependency and temporal regularities of evolving networks via a recurrent learning paradigm. 

Following the concise and effective discrete-time temporal network embedding paradigm, a temporal network is first converted into a series of snapshots over time. In HTGN, we project the nodes into the hyperbolic space, leverage hyperbolic graph neural network (HGNN) to learn the topological dependencies of the nodes at each snapshot, and then employ hyperbolic gated recurrent unit (HGRU) to further capture the temporal dependencies. A temporal network is complex and may have cyclical patterns, and a distant snapshot may be more significant than the closest one~\cite{sankar2020dysat,liu2017global}. Recurrent neural networks (RNNs)~\cite{GRU,LSTM} usually restrict the model to emphasize the most nearby time steps due to its time monotonic assumption. We, therefore, design a wrapped hyperbolic temporal contextual attention (HTA) module that incorporates context from the latest $w$ historical states in hyperbolic space and assigns different weights for both distant and nearby snapshots. % to adaptively assign interpretable weights for previous time steps.
%which incorporates the latest $w$ historical states in hyperbolic space. This is a learning-based module that is flexible and effective to extract the historical information.
On the other hand, temporal coherence serves as a critical signal for sequential learning since a regular temporal graph is usually continuous and smoothly varying. Inspired by the cycle-consistency in video tracking~\cite{wang2019learning,dwibedi2019temporal}, we propose a novel hyperbolic temporal consistency (HTC) component that imposes a constraint on the latent representations of consecutive snapshots, ensuring the stability and generalization for tracking the evolving behaviors. In summary, the main contributions are stated as follows:
\begin{itemize}
\item We propose a novel hyperbolic temporal graph embedding model, named HTGN, to learn temporal regularities, topological dependencies, and implicitly hierarchical organization. To the best of our knowledge, this is the first study on temporal graph embedding built upon a hyperbolic geometry powered by the recurrent learning paradigm.
\item HTGN applies a flexible wrapped hyperbolic temporal contextual attention (HTA) module to effectively extract the diverse scope of historical information.
A hyperbolic temporal consistency (HTC) constraint is further put forward to ensure the stability and generalization of the embeddings. 
%\item We propose a hyperbolic temporal consistency (HTC) constraint for the temporal learning process to ensure the stability and generalization.  
%\item We put forward a wrapped hyperbolic temporal contextual attention (HTA) module in hyperbolic space, which incorporates the latest $w$ historical states in hyperbolic space. This is a learning-based module that is flexible and effective to extract the historical information.
%\item We propose a hyperbolic temporal consistency (HTC) constraint for the temporal learning process to ensure the stability and generalization. 

\item Extensive experimental results on diverse real-world temporal graphs demonstrate the superiority of HTGN as it consistently outperforms the state-of-the-art methods on all the datasets by large margins. The ablation study further gives insights into how each proposed component contributes to the success of the model.
% \textcolor{red}{Besides, HTGN has outstanding performance in extracting hierarchical dependencies, long-term sequence prediction, and low-dimensional representation.}
\end{itemize}

\section{Related works}
Our work mainly relates to representation learning on temporal graph embedding and hyperbolic graph embedding.

\textbf{Temporal graph embedding.} Temporal graphs are mainly defined in two ways: (1) discrete temporal graphs, which are a collection of evolving graph snapshots at multiple discrete time steps; and (2) continuous temporal graphs, which update too frequently to be represented well by snapshots and are instead denoted as graph streams~\cite{TG_survey:foundations}. Snapshot-based methods can be applied to a timestamped graph by creating suitable snapshots, but the converse is infeasible in most situations due to a lack of fine-grained timestamps. Hence, we here mainly focus on representation learning over discrete temporal graphs. For systematic and comprehensive reviews, readers may refer to~\cite{TG_survey:foundations}, and~\cite{TG_survey:Evolutionary}.

The set of approaches most relevant to our work is the recurrent learning scheme that integrates graph neural networks with the recurrent architecture, whereby the former captures graph information and the latter
handles temporal dynamism by maintaining a hidden state to summarize historical snapshots.
% and achieve state-of-the-art results on several temporal graph-related tasks. 
For instance, GCRN~\cite{tgn:GRUGCN} offers two different architectures to capture the temporal and spatial correlations of a dynamic network. The first one is more straightforward and uses a GCN to obtain node embeddings, which are then fed into an LSTM to learn the temporal dynamism. The second is a modified LSTM that takes node features as input but replaces the fully-connected layers therein by graph convolutions. A similar idea is explored in DySAT~\cite{sankar2020dysat}, which instead computes node representations through joint self-attention along the two dimensions of the structural neighborhood and temporal dynamics. VRGNN~\cite{hajiramezanali2019VGRNN} integrates GCRN with VGAE~\cite{kipf2016variational} and each node at each time-step is represented with a distribution; hence, the uncertainty of the latent node representations are also modeled. On the other hand, EvolveGCN~\cite{pareja2020evolvegcn} captures the dynamism of the graph sequence by using an RNN to evolve the GCN parameters rather than the temporal dynamics of the node embeddings. Most of the prevalent methods are built-in Euclidean space which, however, may underemphasize the intrinsic power-law distribution and hierarchical structure.

% reference to Evolegcn/Dysat
%"For a comprehensive survey of representation learning on dynamic networks see~\cite{kazemi2020representation}, and for a survey of dynamic link prediction, including Temporal restricted Boltzmann machines~\cite{TG_survey:Link}"
\textbf{Hyperbolic graph embedding.} Hyperbolic geometry has received increasing attention in machine learning and network science communities due to its attractive properties for modeling data with latent hierarchies. It has been applied to neural networks for problems of computer vision, natural language processing~\cite{nickel2017poincare,gulcehre2019hyperbolicAT,nickel2018learning,sala2018representation}, and graph embedding tasks~\cite{gulcehre2019hyperbolicAT,zhang2019hyperbolic,hgcn2019,liu2019HGNN}. In the graph embedding field, recent works including HGNN~\cite{liu2019HGNN}, HGCN~\cite{hgcn2019}, and HGAT~\cite{zhang2019hyperbolic} generalize the %attention-based 
graph convolution into hyperbolic space (the name of these methods are from corresponding literature) by moving the aggregation operation to the tangent space, where the vector operations can be performed. HGNN~\cite{liu2019HGNN} focuses more on graph classification tasks and provides an extension to dynamic graph embeddings. HGAT~\cite{zhang2019hyperbolic} introduces a hyperbolic attention-based graph convolution using algebraic formalism in gyrovector and focuses on node classification and clustering tasks. HGCN~\cite{hgcn2019} introduces a local aggregation scheme in local tangent space and develops a learnable curvature method for hyperbolic graph embedding. Besides, works in~\cite{gu2019learning,zhu2020gil} propose to learn representations over multiple geometries. The superior performance brought by hyperbolic geometry on static graphs motivates us to explore it on temporal graphs.   
%Compared with the other two counterparts, HGCN~\cite{hgcn2019} additionally provides a local aggregation fashion and a learnable curvature method, which is more generalized.

\section{Preliminary and Background}
In this section, we first present the problem formulation of temporal graph embedding and introduce the widely used graph recurrent neural networks framework. Then, we introduce some fundamentals of hyperbolic geometry.

\subsection{Problem Formulation}
In this work, we focus on discrete-time temporal graph embedding. A discrete-time temporal graph~\cite{TG_survey:Evolutionary,TG_survey:foundations} is composed of a series of snapshots $\{G_{1},...,G_{t},...,G_{T}\}$ sampled from a temporal graph $\mathcal{G}$, where $T$ is the number of snapshots. Each snapshot $G_t=(V_t, A_t)$ contains the current node set $V_t$ and the corresponding adjacency matrix $A_t$. As time evolves, nodes may appear or disappear, and edges can be added or deleted. The graph embedding aims to learn a mapping function that obtains a low-dimensional representation $H_t$ by giving the snapshots till timestamp $t$.   %Discrete-time temporal graph embedding can be viewed as learning a mapping function given the snapshots till timestamp $t$ to obtain a low-dimension representation $Z_t$. 
A general learning framework can be written as:
\begin{equation}
\small
\label{equ:basic_learning_paradigm}
H_t=f_2(f_1(A_t, X_t), H_{t-1}),
\end{equation}
%where $A_t$ is adjacency matrix which is constructed from current edge set $E_t$, 
where $X_t$ is the initial node features or attributes and $H_{t-1}$ is the latest historical state. This learning paradigm is widely used in discrete-time temporal graph embedding~\cite{tgn:GRUGCN,hajiramezanali2019VGRNN,zhao2019tgcn} where $f_1$ is graph neural network, e.g., GCN~\cite{gcn2017} aiming at modeling structural dependencies and $f_2$ is a recurrent network, e.g., GRU~\cite{GRU} to capture the evolving regularities.
% For nodes without attributes, we use one-hot encoding so that the framework can be applied into inductive learning and handling new appearing nodes friendly. Our framework follows the basic architecture of temporal sequence modeling, as shown in the formula~(\ref{equ:basic_framework}).

% \subsection{Graph Recurrent Neural Networks (GRNNs)} 
% To capture both the temporal and structural patterns in a dynamic network, we consider a GRNN framework which contains two basic parts: graph convolution neural network and recurrent neural network. 

% RNN refers to a group spatio-temporal graph neural networks that combine graph neural networks with recurrent neural networks to capture spatial and temporal patterns over nodes of a static graph~\cite{hajiramezanali2019VGRNN}. For instance, GCRN intorudced by~\cite{tgn:GRUGCN} stack graph convolutional networks(GCN) with recurrent long short-term memory(LSTM), and T-GCN~\cite{zhao2019tgcn} combine GCN with gated recurrent unit(GRU) for spatial-temporal traffic forecasting. Given a graph with $G$ with $N$ nodes, whose topology is determined by the adjacency matrix $A\in \mathbb{R}^{N\times N}$, and a sequence of node attributes, GRNN reads the node attributes and updates its hidden state $\mathbf{h}_t\in \mathbb{R}$ at each time step t: 
% \begin{equation}
%   \mathbf{h}_{t}=f_2 \left(f_1(\mathbf{A},\mathbf{x}_{t}),\mathbf{h}_{t-1} \right ),
% \label{equ_formulation}
% \end{equation}
% where $f_1$ is a and $f_2$ can be any recursive network including LSTM or GRU.
\subsection{Hyperbolic Geometry}
\begin{figure}[t]
    \centering
    \includegraphics[width=0.25\textwidth]{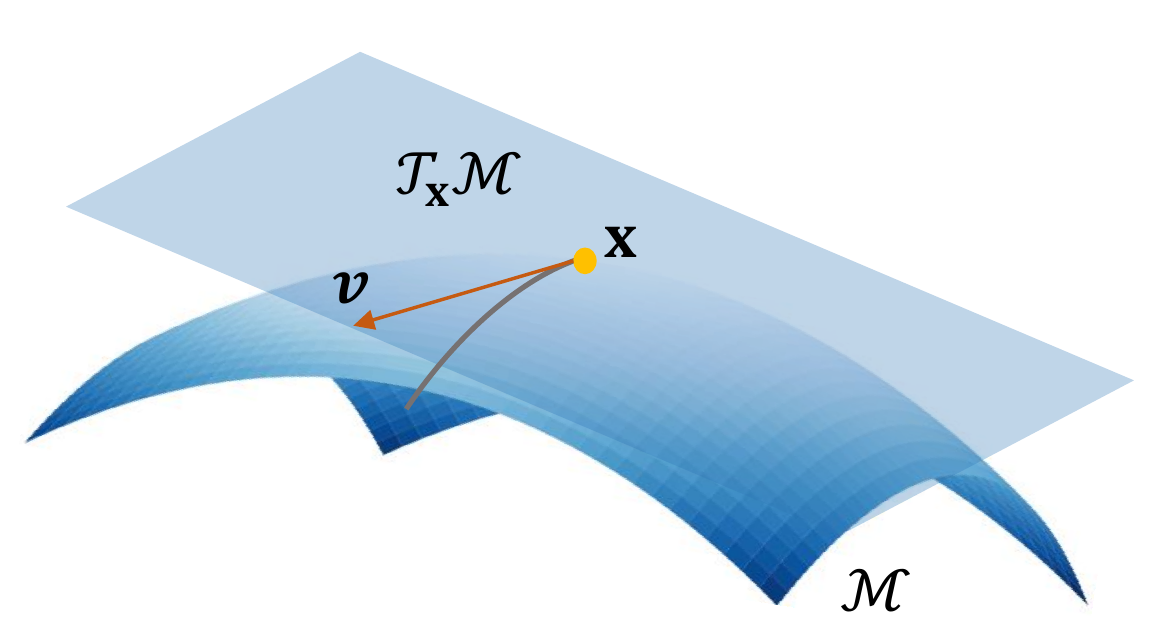}
    \small
    \caption{The tangent space $\mathcal{T}_\mathbf{x}\mathcal{M}$ and a tangent vector $\mathbf{v}$, along the given point $\mathbf{x}$ of a curve traveling through the manifold $\mathcal{M}$. }
    \label{fig:tangent_space}
    \vspace{-10pt}
\end{figure}

A Riemannian manifold $(\mathcal{M},g)$ is a branch of differential geometry that involves a smooth manifold $\mathcal{M}$ with a Riemannian metric $g$. For each point $\mathbf{x}$ in $\mathcal{M}$, there is a tangent space $\mathcal{T}_\mathbf{x}\mathcal{M}$ as the first-order approximation of $\mathcal{M}$ around $\mathbf{x}$, which is a $n$-dimensional vector space (see Figure~\ref{fig:tangent_space}). %In particular, hyperbolic space is a negative curvature Riemannian manifold.

There are multiple equivalent models for hyperbolic space. We here adopt the Poincaré ball model which is a compact representative providing visualizing and interpreting hyperbolic embeddings. The Poincaré ball model with negative curvature $-c$~$(c>0)$ corresponds to
the Riemannian manifold $(\mathbb{H}^{n, c},g_{\mathbb{H}})$, where $\mathbb{H}^{n,c}=\left \{\mathbf{x}\in \mathbb{R}^{n}: c\| \mathbf{x} \|^2 < 1 \right \}$ is an open $n$-dimensional ball. %, then $\frac{1}{\sqrt{c}}$ is the radius of the ball. 
If $c=0$, it degrades to Euclidean space, i.e., $\mathbb{H}^{n,c}=\mathbb{R}^n$. In addition, \cite{HNN} shows how Euclidean and hyperbolic spaces can be continuously deformed into each other and provide a principled manner for basic operations (e.g., addition and multiplication) as well as essential functions (e.g., linear maps and softmax layer) in the context of neural networks and deep learning.

\begin{figure*}[t]
    \centering
    \includegraphics[width=0.9\textwidth]{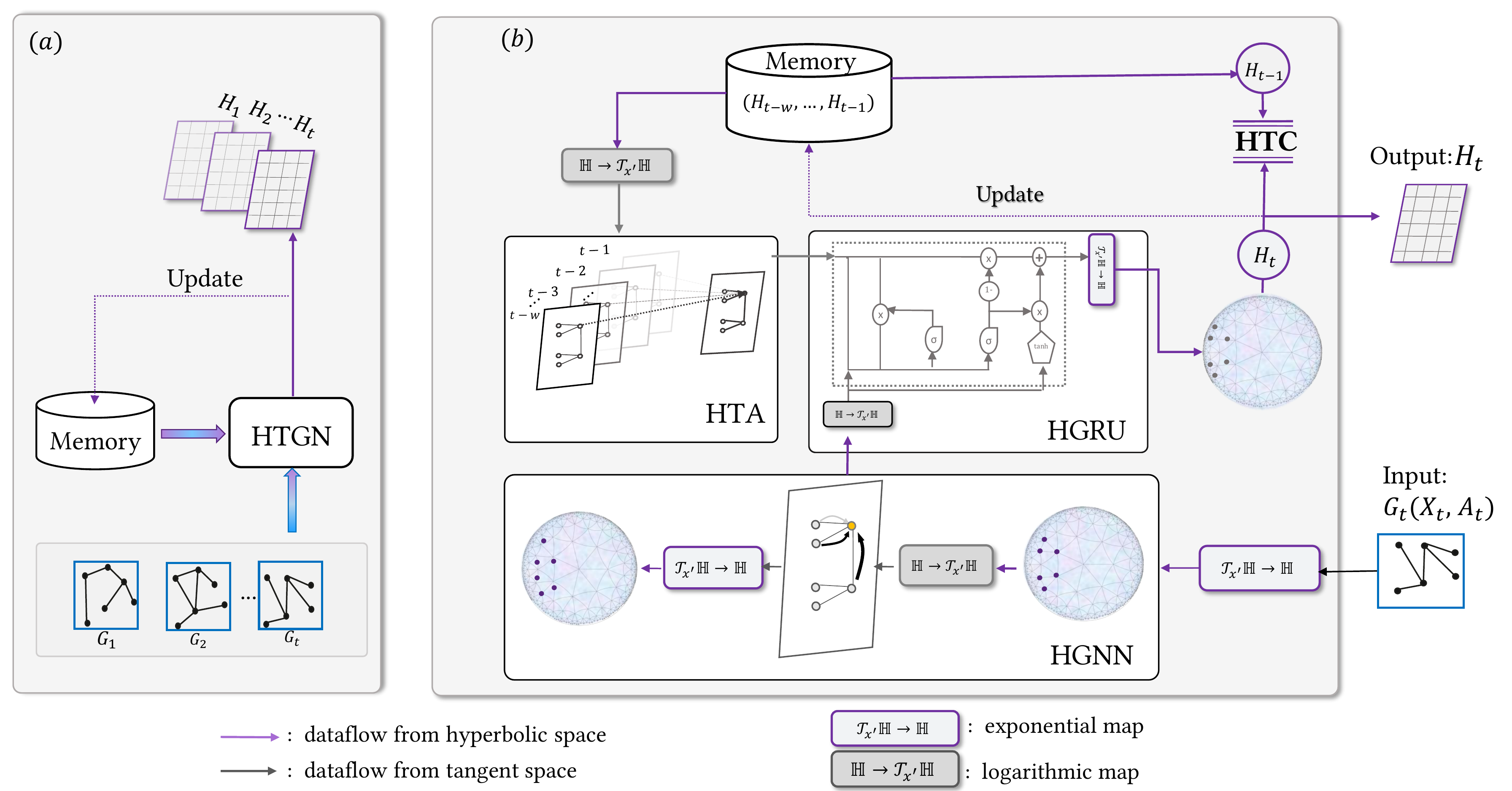}
    \caption{ Schematic of HTGN. Figure~(a) is a sketch of HTGN illustrating the recurrent paradigm and Figure~(b) shows the data flow of an HTGN unit.}
    \label{fig:framework}
    \end{figure*}	

\section{Methodology}
The overall framework of the proposed Hyperbolic Temporal Graph Network (HTGN) is illustrated in Figure~\ref{fig:framework}. As sketched in Figure~\ref{fig:framework}(a), HTGN is a recurrent learning paradigm and falls into the prevalent discrete-time temporal graph architecture formulated by equation~\eqref{equ:basic_learning_paradigm}. An HTGN unit, shown in Figure~\ref{fig:framework}(b), mainly consists of  three components: 
(1) HGNN, the graph neural network to extract topological dependencies in hyperbolic space; 
(2) HTA module, an attention mechanism based on the hyperbolic proximity to obtain the attentive hidden state; 
(3) HGRU, the hyperbolic temporal recurrent module to capture the sequential patterns.
Furthermore, we propose a hyperbolic temporal consistency constraint denoted as HTC to ensure stability and smoothness.
We elaborate on the details of each respective module in the following paragraphs. For the sake of brevity, the timestamp $t$ is omitted in Section~\ref{sec:feature_mapper} and Section~\ref{sec:HGNN}.

% The RNN handles the variable-length sequence by having a recurrent hidden state whose activation at each time is dependent on that of the previous time
\subsection{Feature Map}
\label{sec:feature_mapper}

Before going into the details of each module, we first introduce two bijection operations, the exponential map and the logarithmic map, for mapping between hyperbolic space and tangent space with a local reference point~\cite{liu2019HGNN, hgcn2019}, as presented below. 
% For instance, the input node features are from Euclidean space which are then transferred into  hyperbolic space via exponential mapping. On the other hand, the neural network operations are performed on the tangent space, the logarithmic map is the reverse map that transfers a hyperbolic vector back to the Euclidean space.
%Mapping between tangent space and hyperbolic space is done by the exponential and logarithmic maps, The logarithmic map is the reverse map that maps back to the tangent space at $\mathbf(x)$
\begin{prop}
For $\mathbf{x'}\in \mathbb{H}^{d,c}, \mathbf{a} \in \mathcal{T}_{x'}\mathbb{H}^{d,c}$,  $\mathbf{b}\in \mathbb{H}^{d,c}$, and $\mathbf{a}\neq\mathbf{0}$, $\mathbf{b}\neq \mathbf{x'}$, then the exponential map is formulated as:
\begin{equation}
\small
    \exp_\mathbf{x'}^c(\mathbf{a})= \mathbf{x'}\oplus^c  \left(\mathrm{tanh}(\frac{\sqrt{c}\lambda^c_{\mathbf{x'}} \|\mathbf{a}\|}{2}) \frac{\mathbf{a}}{\sqrt{c}\lVert \mathbf{a}\lVert} \right),
\end{equation}
where $\lambda_\mathbf{x'}^c:=\frac{2}{1-c\|\mathbf{x}'\|^2}$ is conformal factor, and $\oplus$ is the M\"obius addition, for any $\mathbf{u},\mathbf{v}\in \mathbb{H}^{d,c}$:
\begin{equation}
\small
\mathbf{u} \oplus \mathbf{v}:=\frac{\left(1+2 c\langle\mathbf{u}, \mathbf{v}\rangle+c\|\mathbf{v}\|^{2}\right) \mathbf{u}+\left(1-c\|\mathbf{u}\|^{2}\right) \mathbf{v}}{1+2 c\langle\mathbf{u}, \mathbf{v}\rangle+c^{2}\|\mathbf{u}\|^{2}\|\mathbf{v}\|^{2}}.
\end{equation}
The logarithmic map is given by: 
\begin{equation}
\small
\mathrm{log}^c_\mathbf{x'}(\mathbf{b}) :=\frac{2}{\sqrt{c}\lambda^c_{\mathbf{x'}}} \mathrm{arctanh}(\sqrt{c} \lVert -\mathbf{x'} \oplus^c \mathbf{b}\lVert)\frac{-\mathbf{x'} \oplus^c \mathbf{b}}{\lVert -\mathbf{x'} \oplus^c \mathbf{b}\lVert}.
\end{equation}
\end{prop}
\noindent
Note that $\mathbf{x'}$ is a local reference point, we use the origin point $\mathbf{0}$ in our work.
% Given a Euclidean space vector $\mathbf{x}_i^E \in \mathbb{R}^d$, we regard it as the point in the tangent space $\mathcal{T}_{\mathbf{x}'}\mathbb{H}^{d,c}$ with the reference point $\mathbf{x}'\in \mathbb{H}^{d,c}$ (Similar to ~\cite{liu2019HGNN,hgcn2019},  we here also use original point as the reference point), and then the exponential map is defined as:

% When the reference point is origin, i.e., $\mathbf{x'}=0$, the exponential map is formulated as:
% \begin{equation}
%     \exp_0^c(\mathbf{v}) = \tanh(\sqrt{c}\|v\|)\frac{v}{\sqrt{c}\|v\|}
% \end{equation}
%In the following operations, we need to utilize 
% when $\mathbf{x}=0$, we have:
% \begin{equation}
%     \log_0^c(x')=\mathrm{arctanh}(\sqrt{c}\|y\|)\frac{y}{\|y\|}
% \end{equation}

\subsection{Hyperbolic Graph Neural Network (HGNN)}
\label{sec:HGNN}
%HGNN is employed to learn topological dependencies in the temporal graph leveraging promising properties of hyperbolic geometry in each timestamp. Analogous to GNN, a HGNN layer also includes three key operations: hyperbolic transformation, hyperbolic aggregation and hyperbolic activation. Given $\mathbf{x}_i^\mathcal{H}$ in a graph, the update rule for one HGNN layer is expressed as:
HGNN is employed to learn topological dependencies in the temporal graph leveraging promising properties of hyperbolic geometry. Analogous to GNN, an HGNN layer also includes three key operations: hyperbolic transformation, hyperbolic aggregation, and hyperbolic activation. Given a Euclidean space vector $\mathbf{x}_i^E \in \mathbb{R}^d$, we regard it as the point in the tangent space $\mathcal{T}_{\mathbf{x}'}\mathbb{H}^{d,c}$ with the reference point $\mathbf{x}'\in \mathbb{H}^{d,c}$ and use the exponential map to project it into hyperbolic space, obtaining $\mathbf{x}_i^\mathcal{H}\in\mathbb{H}^{d,c}$, 
\begin{equation}
\small
    \mathbf{x}_i^\mathcal{H} = \exp_{\mathbf{x^\prime}}^c(\mathbf{x}_i^E).
\end{equation}
Then the update rule for one HGNN layer is expressed as:
\begin{subequations}
\small
\label{equ:HGNN}
\begin{align}
   \mathbf{m}_i^{\mathcal{H}}&=W\otimes^c\mathbf{x}_i^{\mathcal{H}}\oplus^c \mathbf{b},\tag{\ref{equ:HGNN}a}\label{equ:HGNNa}\\
   \tilde{\mathbf{m}}_i^{\mathcal{H}}&=\exp_{\mathbf{x'}}^c(\sum_{j \in \mathcal{N}(i)} \alpha_{ij}\log_\mathbf{\mathbf{x}'}^c(\mathbf{m}_i^{\mathcal{H}}))\tag{\ref{equ:HGNN}b},\label{equ:HGNNb}\\
   \tilde{\mathbf{x}}_i^{\mathcal{H}}&= \exp_\mathbf{x'}^c(\sigma({\log_{\mathbf{x'}}^c}(\tilde{\mathbf{m}}_i^{\mathcal{H}})).\tag{\ref{equ:HGNN}c}\label{equ:HGNNc}
\end{align}
\end{subequations}
\noindent
\textbf{Hyperbolic Linear Transformation (\ref{equ:HGNNa}).} 
%Linear transformation contains both vector multiplication and bias addition.directly applied in hyperbolic space as it fail to meet the requirements of permutation invariant since the basic operations in hyperbolic space, such as M\"obius addition cannot satisfy commutative and associative properties. Therefore, for vector multiplication, we use logarithmic mapping to project hyperbolic vector in tangent space and apply multiplication, which is given by:
The hyperbolic linear transformation contains both vector multiplication and bias addition which can not be directly applied since the operations in hyperbolic space fail to meet the permutation invariant requirements. % since the basic operations such as M\"obius addition cannot satisfy commutative and associative properties. 
Therefore, for vector multiplication, we first project the hyperbolic vector to the tangent space and apply the operation, which is given by:
\begin{equation}
\small
    W\otimes^c
    \mathbf{x}_i^\mathcal{H}:= \exp_\mathbf{x'}^c(W\log_\mathbf{x'}^c(\mathbf{x}_i^{\mathcal{H}})).
\end{equation}
For bias addition, we transport a bias $\mathbf{b}$ located at $\mathcal{T}_\mathbf{o}\mathbb{H}$ to the position $\mathcal{T}_\mathbf{x}\mathbb{H}$ in parallel. Then, we use $\exp_{\mathbf{x}}^c$ to map it back to hyperbolic space:
\begin{equation}
    \mathbf{x}^\mathcal{H}\oplus^c \mathbf{b}:=\exp_{\mathbf{x}}^c(P_\mathbf{o\to \mathbf{x}}(\mathbf{b})).
\end{equation}
\noindent
\textbf{Hyperbolic Aggregation (\ref{equ:HGNNb}).} The (weighted) mean operation is necessary to perform aggregation in an Euclidean graph neural network. An analog of mean aggregation in hyperbolic space is the  Fr\'echet mean~ \cite{frechet1948elements}, however, it is difficult to apply as it lacks a closed-form to compute the derivative easily~\cite{bacak2014computing}.
\input{each_parts/aggregation}
\noindent
\textbf{Hyperbolic Activation (\ref{equ:HGNNc})}. As given in equation~\eqref{equ:HGNNc}, the hyperbolic activation is achieved by applying logarithmic and exponential mapping. Noted that the two curvatures can be different.   %Similarly, we use logarithmic and exponential mapping to achieve hyperbolic activation as given by equation~\eqref{equ:HGNNc}.

\subsection{Hyperbolic Temporal Attention (HTA)}
Historical information plays an indispensable role in temporal graph modeling since it facilitates the model to learn the evolving patterns and regularities. %\textcolor{red}{Taking social networks as examples, a person's online shopping, forwarding posts on social networks, interacting with friends in communication networks are not only related to the past one hour or day but more likely related to recent activities.}
Although the latest hidden state $H_{t-1}$ obtained by the recurrent neural network already carries historical information before time $t$, some discriminative contents may still be under-emphasized due to the monotonic mechanism of RNNs that temporal dependencies are decreased along the time span~\cite{liu2017global}.% This  motivates us to expand the scope of historical information usage in the process of temporal graph modeling.  %the catastrophic forgetting of RNNs. This  motivates us to expand the scope of historical information usage in the process of temporal graph modeling.

Inspired by~\cite{HCA}, our proposed HTA generalizes $H_{t-1}$ to the latest $w$ snapshots $\{H_{t-w},\cdots, H_{t-1}\}$, attending on multiple historical latent states to get a more informative hidden state. The procedure is illustrated in Algorithm~\ref{alg:whta}.
Specifically, we first project $w$ historical states in the state memory bank into the tangent space and concatenate them together. Then, the learnable weight matrix $Q$ and vector $\mathbf{r}$ are utilized to extract contextual information, where $Q$ weights the node importance in each historical state and $\mathbf{r}$ determines the weights across time windows.%, and the Softmax is applied to get the attention scores. %Next, we compute an attentive hidden state by the multiplication of attention scores and original hidden states. Finally, the node state is projected back to hyperbolic space for the next module.
%Specifically, we first project $w$ historical states in the state memory bank into tangent space and concatenate them together. We further utilize the learnable weight matrix $Q$ and vector $\mathbf{r}$ to extract important information, where $Q$ weights the node importance in one historical state and $r$ determines the weights across time windows. The softmax function is further  to get the attention scores along the time dimension for each node. Next, we compute an attentive hidden state by the multiplication of attention scores and original hidden states. Last, we map the node state back to hyperbolic space. 
\begin{algorithm}[t] 
\small
\begin{flushleft}
 \textbf{Input:} $\{H_{\tau}^\mathcal{H}\}_{\tau=t-w}^{\tau=t-1}$ \\
 \textbf{Output:} ${\tilde{H}_{t-1}}$
\end{flushleft}
 \begin{algorithmic}[1]
   \STATE \textbf{for} $\tau=t-w$ to $t-1$ \textbf{do}
    \INDSTATE  $H_{\tau}^E = \log_\mathbf{x'}^c(H_{\tau}^\mathcal{H})$ 
    \STATE \textbf{end for}
    \STATE $M = [H_{t-w}^E\|,...,\| H_{t-1}^E]$
    \STATE $A_{tt} = r^T\tanh(Q M)$
    \STATE $\tilde{A}_{tt[i,:]} = softmax(A_{tt[i,:]})$
    \STATE $H_{t-1}^E= \tilde{A}_{tt}M$
    \STATE \textbf{return} ${\tilde{H}_{t-1}^\mathcal{H}}=\exp_\mathbf{x'}^c({H}_{t-1}^E)$
% \end{aligned}
% \end{equation} 
  \end{algorithmic} 
  \caption{HTA learning procedure}
  \label{alg:whta}
\end{algorithm}

% \begin{algorithm}[H]
% \begin{flushleft}
% \textbf{Input:} Hyperbolic $w$ historical hidden states $\{H_{t-w} ,..., H_{t-1}\}$  \\
% \textbf{Parameter:}\\
% \textbf{Output:} An attentive state $\tilde{H}$ located on tangent space $\mathcal{T}_\mathbf{x'}\mathbb{H}$.\\
% \SetAlgoLined
% (1) \textbf{for} $\tau=T-w$ to $t-1$ \textbf{do}:\\
% \INDSTATE \quad (i) project $H_\tau$ into tangent space $\mathcal{T}_\mathbf{x'}\mathbb{H}$:\\
% \begin{equation}
%     H_{\tau}^E = \log_\mathbf{x'}^c(H_{\tau}) 
% \end{equation} 
% \INDSTATE \quad (ii) concatenate all projected states\\
% \begin{equation}
%     H_w^E = [H_{t-w}^E\|,...,\| H_{t-1}^E]
% \end{equation} 
% \INDSTATE \quad (iii) compute the temporal contextual attention scores: \\
% \begin{equation}
% \begin{aligned}
%     e_i &= r^T \tanh(Q(H_w^E)^T) \\
%     \alpha_i &= softmax(e_i) \\
% \end{aligned}
% \end{equation} 
% (2) Compute attentive hidden state over all nodes;
% \begin{equation}
%     \tilde{H}_w= A H_w ^T
% \end{equation}
% where $A$ is attention matrix and each row is $\alpha_i$ ($i\in [1, N]$)
% \caption{Hyperbolic Temporal Attention}
% \label{alg:whta}
% \end{flushleft}
% \end{algorithm}

\subsection{Hyperbolic Gated Recurrent Unit (HGRU)}
GRU~\cite{GRU}, a variant of LSTM~\cite{LSTM}, is used in this work to incorporate the current and historical node states. Similar to LSTM, the GRU adopts gating units to modulate the flow of information but gets rid of the separate memory cell. 
%GRU has got rid of the cell state of LSTM and uses the hidden state to transfer information. 
%Benefiting from its parsimonious structure, the performance of a GRU is on a par with LSTM while being more computationally efficient. 
Note that, our HGRU\footnote{A HyperGRU defined by~\citeauthor{HNN}~\cite{HNN} is also applicable in our framework. However, we experimentally found that the proposed method built in the tangent space $\mathcal{T}_{\mathbf{x'}}\mathcal{H}$ obtains similar performance but is more efficient for large-scale data.} is performed in the tangent space due to its computational efficiency.

HGRU receives the sequential input $\tilde{X}_t^\mathcal{H}$ from HGNN and the attentive hidden state $\tilde{H}_{t-1}^\mathcal{H}$ obtained from HTA as the input, and we denote $H_t^\mathcal{H}$ as the output. The dataflow in the HGRU unit is characterized by the following equations: 

%Let current state matrix $\tilde{X}_t^\mathcal{H}$ and historical state $\tilde{H}_{t-1}^\mathcal{H}$ be the outcomes from HGNN and HTA modules, respectively. It should be noted that the GRU\footnote{A GRU built in hyperbolic space is also applicable in our framework~\cite{HNN}. However, we experimentally found that the GRU building in the tangent space obtains the similar performance but is more efficient for large-scale data.} this is performed in the tangent space, so logarithmic mapping is required (equations (\ref{equ:GRU}a, \ref{equ:GRU}b)). Then, we feed the states into the GRU directly (equations (\ref{equ:GRU}c-\ref{equ:GRU}f)). After that, we map it back to hyperbolic space (equation (\ref{equ:GRU}g)). 
\begin{subequations}
\small
\label{equ:GRU}
\begin{align}
&X_t^E = \log_{\mathbf{x'}}^c(\tilde{X}_t^\mathcal{H})\tag{\ref{equ:GRU}a},\label{equ:GRUa}\\
&H_{t-1}^E = \log_{\mathbf{x'}}^c(\tilde{H}_{t-1}^\mathcal{H})\tag{\ref{equ:GRU}b},\label{equ:GRUb}\\
& P_t^E = \sigma(W_z X_t^E + U_z H_{t-1}^E) \tag{\ref{equ:GRU}c}\label{equ:GRUc}\\
& R_t^E = \sigma(W_r X_t^E + U_r H_{t-1}^E), \tag{\ref{equ:GRU}d}\label{equ:GRUd}\\
& \tilde{H}_t^E = \tanh(W_h X_t^E + U_h (R_t\odot H_{t-1}^E)), \tag{\ref{equ:GRU}e}\label{equ:GRUe}\\
& H_t^E = (1-P_t^E) \odot \tilde{H}_t^E + P_t^E \odot H_{t-1}^E, \tag{\ref{equ:GRU}f}\label{equ:GRUf}\\
& H_t^\mathcal{H} = \exp_{\mathbf{x'}}^c(H_t^E). \tag{\ref{equ:GRU}g}\label{equ:GRUg}
\end{align}
\end{subequations}
where $W_z, W_r, W_h, U_z, U_r, U_h$ are the trainable weight matrices, $P_t^E$ is the update gate to control the output and $R_t^E$ is the reset gate to balance the input and memory. As the GRU is built in the tangent space, logarithmic maps are needed (equations \eqref{equ:GRUa}, \eqref{equ:GRUb}). Then, we feed the states into the GRU (equations \eqref{equ:GRUc} to \eqref{equ:GRUf}) and map the hidden state back to hyperbolic space (equation \eqref{equ:GRUg}). As we can see, the final $H_t^{\mathcal{H}}$ fuses structural, content, and temporal information.

\subsection{Proposed Learning Algorithm}
%\subsubsection{ The Objective Function}
Uniting the above modules, we have the overall learning procedure as summarized in Algorithm~\ref{alg:HTGN_procedure}. In line 5, we also consider include the historical state as the input as ~\cite{hajiramezanali2019VGRNN} and for brevity, we ignore it here. %For a given snapshot, the node features are interpreted into hyperbolic space and HGNN is then applied to extract the topology dependency. Parallelly, the attention is performed on the historical hidden states.  %\textcolor{red}{TBA}
\begin{algorithm}[t]
\small
\begin{flushleft}
\textbf{Input:} Node interaction stream $\{A_t\}_{t=1}^{t=T}$ and attributes $\{X_t^E\}_{t=1}^{t=T}$.\\%, and learnable curvature $c$.\\
\textbf{Output:} $H_T^\mathcal{H}, c$.
\end{flushleft}
  \begin{algorithmic}[1]
   \STATE Initialize $w\times\{H_0^\mathcal{H}\}$ and curvature $c$
    \STATE \textbf{repeat}
    \INDSTATE \textbf{for} $t=1$ to $T$ \textbf{do}
    \INDSTATE[2] $X_t^{\mathcal{H}}=\exp_{\mathbf{x'}}^c(X_{t}^E)$
    % \INDSTATE[2] $\tilde{X}_t^{\mathcal{H}}$ = $\mathbf{HGNN}(\exp_\mathbf{x'}^c[\log_\mathbf{x'}^c(X_t^{\mathcal{H}})||\log_\mathbf{x'}^c(H_{t-1}^\mathcal{H})])$
    \INDSTATE[2] $\tilde{X}_t^{\mathcal{H}}$ = $\mathbf{HGNN}(X_t^{\mathcal{H}})$
    \INDSTATE[2] $\tilde{H}_{t-1}^{\mathcal{H}}$ = $\mathbf{HTA}(H_{t-w};...;H_{t-1})$
     \INDSTATE[2]
     $H_t^\mathcal{H}=\mathbf{HGRU}(\tilde{X}_t^{\mathcal{H}}, \tilde{H}_{t-1}^\mathcal{H})$
     %$H_t^\mathcal{H}=\exp_{\mathbf{x'}}^c\left(\mathbf{GRU}(\log_{\mathbf{x'}}^c(\tilde{X}_t^{\mathcal{H}}), \log_{\mathbf{x'}}^c(\tilde{H}_{t-1}^\mathcal{H}))\right)$
     \INDSTATE[2] $\mathcal{L}_t = \mathcal{L}_{t,c}+\lambda\mathcal{L}_{t,r}$
     \INDSTATE[2] Minimize $\mathcal{L}_t$
     \INDSTATE[2] Update state memory bank
     \INDSTATE \textbf{end for}
    \STATE \textbf{until} Convergence
    \STATE \textbf{return} $H_T^\mathcal{H}, c$
  \end{algorithmic} 
  \caption{The learning procedure of HTGN}
  \label{alg:HTGN_procedure}
\end{algorithm}
Note that we design the objective function $\mathcal{L}_t$ from two aspects: temporal evolution and topological learning, corresponding to the following hyperbolic temporal consistency loss and hyperbolic homophily loss. %It is notable that both of them are unsupervised.
\subsubsection{Hyperbolic Temporal Consistency Loss}
In terms of the time perspective, intuitively, the embedding position in the latent space changes gradually over time, which ensures stability and generalization. We thus pose a hyperbolic temporal consistency constraint $\mathcal{L}_{t,c}$ on two consecutive snapshots $(G_t, G_{t-1})$, to ensure the representation a certain temporal smoothness and long-term prediction ability, which is defined as: 
\begin{equation}
\small
    \mathcal{L}_{t,c} = \frac{1}{N}\sum_{i=1}^N d^\mathcal{H}(\mathbf{x}_{t,i}^\mathcal{H}, \mathbf{x}_{(t-1),i}^\mathcal{H}),
\end{equation}
where the subscript $t$ denotes the loss is with respect to time snapshot $t$, and $d^\mathcal{H}$ is the distance of two points $\mathbf{u},\mathbf{v}\in \mathbb{H}^{d'}$:
\begin{equation}
\small
 d^\mathcal{H}(\mathbf{u},\mathbf{v}) = \frac{2\mathrm{arctanh}\left (\sqrt{c}\lVert -\mathbf{u}\oplus^c\mathbf{v}\lVert \right) }{\sqrt{c}}.
\label{equ:poincare}
\end{equation}

\subsubsection{Hyperbolic Homophily Loss}
\label{section:loss}
Graph homophily that linked nodes often belong to the same class or have similar attributes is a property shared by many real-world networks. The hyperbolic homophily loss $\mathcal{L}_{t,r}$ aims to maximize the probability of linked nodes through the hyperbolic feature and minimize the probability of no interconnected nodes. $\mathcal{L}_{t,r}$ is based on cross-entropy where the probability is inferred by the Fermi-Dirac function~\cite{2010hyperbolic,hgcn2019} which is formulated as: 
\begin{equation}
\small
    p_f(\mathbf{x}_{i}^\mathcal{H}, \mathbf{x}_j^\mathcal{H}):= [\exp((d^\mathcal{H}(\mathbf{x}_i^\mathcal{H}, \mathbf{x}_j^\mathcal{H})-r)/s)]^{-1},
    \label{fermi_dirac}
\end{equation}
where $r$ is the radius of $\mathbf{x}_i^\mathcal{H}$, so points within that radius may have an edge with $\mathbf{x}_i^\mathcal{H}$, and $s$ specifies the steepness of the logical function. Then, $\mathcal{L}_{t,r}$ is given by:
\begin{equation}
\small
\mathcal{L}_{t, r}=\frac{1}{E_1}\sum_{e_{ij}\in \mathcal{E}_t}-\log(p_f(\mathbf{x}_{t,i}^\mathcal{H}, \mathbf{x}_{t,j}^\mathcal{H})) - \frac{1}{E_2}\sum_{e_{i'j'}\notin \mathcal{E}_t}(1-\log(p_f(\mathbf{x}_{t,i'}^\mathcal{H}, \mathbf{x}_{t,j'}^\mathcal{H})).
\end{equation}
To efficiently compute the loss and gradient, we sample the same number of negative edges as there are positive edges for each timestamp, i.e., $E_1=E_2=|\mathcal{E}_t|$.

\subsubsection{The Unified Model}
As temporal consistency and homophily regularity mutually drive the evolution of the temporal graphs, we set the final loss function as:  
\begin{equation}
\small
    \mathcal{L}_t = \mathcal{L}_{t,r} + \lambda\mathcal{L}_{t,c},
\end{equation}
where $\lambda \in[0,1]$ is the hyper-parameter to balance the temporal smoothness and homophily regularity. 

\input{each_parts/proposition}

%\subsection{HTGN}
%\subsubsection{Learning Procedure}
%Uniting the above modules, we have the overall learning procedure as summarized in Algorithm~\ref{alg:HTGN_procedure}. In detail, for a given snapshot, the node features are interpreted into hyperbolic space and HGNN is then applied to extract the topology dependency. Parallelly, the Note that \textcolor{red}{TBA}
%In detail, for a snapshot, we first interpret the current node feature $X_t^E\in \mathbb{R}^{n\times d}$ into hyperbolic space $X_t^\mathcal{H}\in \mathbb{H}^{n\times d}$, and then utilize HGNN to extract the topological dependencies, receiving $\tilde{X}_t^\mathcal{H}\in \mathbb{H}^{n\times d'}$. At the same time, we incorporate latest $w$ historical states (hyperbolic tensor) using the proposed HTA module and obtain an attentive representation $\tilde{H}_{t-1}^{\mathcal{H}}\in \mathbb{H}^{n\times d'}$; Then, We feed $\tilde{X}_t^\mathcal{H}$ and $\tilde{H}_{t-1}^{\mathcal{H}}$ into the HGRU, getting $H_t^\mathcal{H}\in \mathbb{H}^{n\times d'}$. It is worth noting that the GRU is carried out on the tangent plane, which retains the characteristics of the hyperbolic geometry. After that, we deploy the loss function defined in section~\ref{section:loss} to optimize the model. Last but not least, we update the state memory bank using $H_t^\mathcal{H}$. The output is the node embedding $H_t^\mathcal{H}$ in hyperbolic space and the corresponding curvature $c$.  

\begin{table}[t]
\centering
\caption{Complexity analysis.}
\resizebox{0.25\textwidth}{!}{%
\small
\begin{tabular}{@{}c|cccccc@{}}
\toprule
\textbf{Components}        & Time Complexity \\ \midrule \midrule
\textbf{HTA} & $O(Nwd'^2)$      \\
\textbf{HTC}&  $O(Nd)$   \\
1-layer \textbf{HGNN}  & $O(Ndd' + d'|\mathcal{E}_t|)$    \\
\bottomrule
\end{tabular}%
}
\label{tab:complexity}
\end{table}

\subsubsection{Complexity Analysis}
\label{sec:complexity}
We analyze the time complexity of the main components of the proposed HTGN model in each timestamp and present a summary in Table~\ref{tab:complexity}, where $N$ and $|\mathcal{E}_t|$ are the number of nodes and edges, $d$ and $d'$ are respectively the dimensions of input feature and output feature, and $w$ denotes state memory length. %For HGNN, the time complexity is $O(ndd' + d'|\mathcal{E}|)$, where $n$ and $|\mathcal{E}|$ is the number of nodes and edges, $d$ and $d'$ is the dimension of input feature and output feature. For HTA module, the time complexity is $O(nwd')$ for the $w$ length state memory.  For HTC module, the time complexity is $O(nd)$.
Note that the above modules can be paralleled across all nodes and are computationally efficient. Furthermore, as we use a constant memory state bank, the extra storage cost is negligible. Numerical analysis on the scalability is presented in Section~\ref{sec:scalability}.

\section{Experiments and analysis}

% \usepackage{booktabs}
% \usepackage{graphicx}

% \begin{table}[!hb]
% \caption{Summary statistics of the benchmark datasets. For each dataset, we report the number of nodes, links, and time steps. Link counts include the number of snapshots containing each link, e.g., a link that occurs in two snapshots, is counted twice.} %dysat
% \label{tab:graph_summary}
% \vskip 0.15in
% \begin{center}
% \begin{small}
% \begin{sc}
% \begin{tabular}{lcccccc}
% \toprule
% {\upshape Dataset}  &Enron & DBLP & FB & IMDB-B & IMDB-M & COLLAB \\
% \midrule
% %{\upshape Source}&Bio&Bio&Bio&Social&Social&Social\\ 
% {\upshape $\#_{Nodes}$} &143 &1,809 &2 &2 &2 &3 \\
% {\upshape $\#_{Links}$}&2,247&16,822&39.06&19.77&13&74.49\\
% {\upshape $\#_{Time steps}$}& 12&13&72.81&193.06&131.87&4914.99\\
% \bottomrule
% \end{tabular}
% \end{sc}
% \end{small}
% \end{center}
% \vskip -0.1in
% \end{table}

In this section, we conduct extensive experiments with the aim of answering the following research questions (\textbf{RQs}): %\textcolor{red}{we evaluate the proposed HTGN using diverse networks comparing various competing temporal graph models, with the aim of answering the following research questions}:
\begin{itemize}
    \item \textbf{RQ1}. How does HTGN perform?% compared to state-of-the-arts?
    \item \textbf{RQ2}. What does each component of HTGN bring? %of HTGN contribute?% to performance?
\end{itemize}
\subsection{Experimental Setup}
\subsubsection{Datasets}
To verify the generality of our proposed method, we choose a diverse set of networks for evaluation, including disease-spreading networks, DISEASE; academic co-author networks, HepPh and COLAB; social networks, FB; email communication networks, Enron; and Internet router network, AS733, as recorded in Table~\ref{tab:dataset_statistics}. Notable, DISEASE is a synthetic dataset based on the SIR spreading model~\cite{bjornstad2002dynamics}, which is also feasible for COVID-19 path prediction. At the same time, we list the Gromov’s hyperbolicity $\delta$~\cite{Gromov_hyperbolicity1,Gromov_hyperbolicity2} and the average density $\rho$. Gromov’s hyperbolicity is a notion from graph theory and measures the ``tree-likeness'' of metric spaces. The lower hyperbolicity, the more tree-like, with $\delta=0$ denoting a pure tree. The average density is defined as the ratio of the number of edges and all possible edges, describing how dense a graph is. In these datasets, HepPh and COLAB are relatively dense and FB is highly sparse. More details about the datasets are presented in Appendix~\ref{appendix:data_processing}.

\begin{table}[t]
\centering
\caption{Dataset statistics.}
\label{tab:dataset_statistics}
\resizebox{0.475\textwidth}{!}{%
\begin{threeparttable}
    \begin{tabular}{@{}lcccccc@{}}
    \toprule
    \textbf{Datasets}      & \textbf{DISEASE} & \textbf{HepPH} & \textbf{FB} & \textbf{AS733} & \textbf{Enron} & \textbf{COLAB} \\ \midrule \midrule
    \textbf{\#Snapshots}   & 7                & 36             & 36          & 30             & 11             & 10             \\
    \textbf{\#Test k}      & 3                & 6              & 3           & 10             & 3              & 3              \\
    \textbf{\#Nodes}       & 2665             & 15,330         & 45,435      & 6,628          & 184            & 315            \\
    \textbf{\#Total Edges} & 2664             & 976,097        & 180,011     & 13,512         & 790            & 943            \\
    \textbf{Density} $\rho~(0.01)^{*}$           & 0.41             & 1.37           & 0.04        & 0.2            & 3.37           & 0.94           \\
    \textbf{Hyperbolicity} $\delta^{\star}$         & 0.0                &   1.0             & 2.0             & 1.5               & 1.5            & 2.0              \\ \bottomrule
\end{tabular}%
\begin{tablenotes}
        %\footnotesize
        \small
        \item[*] $0.01$ denotes the value is in units of 0.01.
        \item[$\star$] The smaller $\delta$  indicates the dataset has a more evident hierarchical structure.
    \end{tablenotes}
\end{threeparttable}
}
\vspace{-10pt}
\end{table}

\input{each_parts/auc_ap_table}

\subsubsection{Baselines}

We compare the performance of our proposed model against a diverse set of competing graph embedding methods. The first two are advanced static network embedding models: GAE and VGAE\footnote{\url{https://github.com/tkipf/gae}}. \textbf{GAE} is simply composed of two-layer graph convolutions; \textbf{VGAE} additionally introduces variantial variables. Compared with the graph models tailored for temporal graphs, the static models ignore the temporal regularity. We use all the edges in the training shots for training and the remaining as the test set. %In order to maintain evaluation consistency, we use micro-AUC/AP as the final results as dynamic models.
% \textcolor{red}{For a fair comparison, we use micro-AUC/AP as the evaluation metric.} 
We moreover compare with \textbf{GRUGCN}, conceptually the same version as in \cite{tgn:GRUGCN} and also a basic architecture of temporal graph embedding model, to show the effectiveness of Hyperbolic geometry and our proposed HTA module. More importantly, we also conduct experiments on several state-of-art temporal graph embedding models: \textbf{EvolveGCN}\footnote{There are two versions of the EvolveGCN: EvolveGCN-O and EvolveGCN-H. We test both and report the best result.}~\cite{pareja2020evolvegcn}, \textbf{DySAT}~\cite{sankar2020dysat}, and \textbf{VGRNN}~\cite{hajiramezanali2019VGRNN} to further demonstrate the superiority of the proposed HTGN. %for temporal graph embedding. 
%\textbf{EvolveGCN}\footnote{\url{https://github.com/IBM/EvolveGCN}} uses RNN to evolve GCN parameters, aiming at preserving temporal patterns in its weights. There are two versions of the EvolveGCN method:
%EvolveGCN-O and EvolveGCN-H, and we report the best results of them. \textbf{DySAT}\footnote{\url{https://github.com/aravindsankar28/DySAT}} utilizes temporal attention and structure attention along temporal and structure dimensions. \textbf{VGRNN}\footnote{\url{https://github.com/VGraphRNN/VGRNN}} introduces additional
%latent random variables and also leverages GRNN framework to model temporal graphs.

\subsubsection{Evaluation Tasks and Metrics}
We obtain node representations from HTGN which can be applied to various downstream tasks. In temporal graph embedding, link prediction is widely used for evaluation, as the addition or removal of edges over time leads to the network evolution. Here, we use the Fermi-Dirac function defined in equation~\eqref{fermi_dirac} to predict links between two nodes. Similar to VGRNN~\cite{hajiramezanali2019VGRNN}, we evaluate our proposed models on two different dynamic link prediction tasks: \textit{temporal link prediction} and \textit{temporal new link prediction}. More specifically, given partially observed snapshots of a temporal graph $\mathcal{G}=\left \{ G_{1},...,G_{t}\right\}$, \textit{dynamic link prediction} task is defined to predict the link in the next snapshots $G_{t+1}$ or next multi-step snapshots and \textit{dynamic new link prediction} task is to predict new links in $G_{t+1}$ that are not in $G_{t}$.

Following the same setting as in VGRNN~\cite{hajiramezanali2019VGRNN}, we choose the last $k$ snapshots as the test set and the rest of the snapshots as the training set. To thoroughly verify the effectiveness of the model, we select different lengths for testing and the corresponding $k$ values are listed in Table~\ref{tab:dataset_statistics}. 
%As for the metric, w
We test the models regarding their ability of correctly classifying true and false edges by computing average precision (AP) and area under the ROC curve (AUC) scores. We assume all known edges in the test snapshots as true and sample the same number of non-links as false. Note that we uniformly train both the baselines and HTGN by using early stopping based on the performance of the training set.  %\textcolor{red}{The reported results are based on the average loss on the training snapshots and early stopping techniques.}  %For more details about experiment settings, please see the appendix.

%Following the experimental settings in VGRNN~\cite{hajiramezanali2019VGRNN}, we choose the last $k$ snapshots as test set and the rest snapshots as training set. To fully verify the effectiveness of our model, we select different lengths for testing where the $k$ is illustrated in Table 1. As for the metric, we test different models based on their ability to correctly classify true and false edges by computing average precision (AP) and area under the ROC curve (AUC) scores. We assume all edges in the test snapshots as true edges and sample the same number of non-links as false edges. 

\subsection{Experimental Results (RQ1)}
The code of HTGN is publicly available here.\footnote{\url{https://github.com/marlin-codes/HTGN}}
We repeat each experiment five times and report the average value with the standard deviation on the test sets in Table~\ref{tab:result_link_prediction} and Table~\ref{tab:result_new_link_prediction}, where the best results are in bold and the second-best results are in italics for each dataset. It is observed that HTGN consistently and significantly outperforms the competing methods of both tasks across all six datasets, demonstrating the effectiveness of the proposed method. On the other hand, the runners-up go to the other temporal graph embedding methods, which confirms the importance of temporal regularity in temporal graph modeling. In the following, we discuss the results on link prediction and new link prediction, respectively.
%in both of the tasks: link prediction and new link prediction for all of the dataset, which demonstrates the effectiveness of the proposed models.

\subsubsection{Link Prediction} Table~\ref{tab:result_link_prediction} shows the experiments on the link prediction task. In summary, HTGN outperforms the competing methods significantly considering both AUC and AP scores, which shows that our proposed models have better generalization ability to capture temporal trends. For instance, HTGN achieves an average gain of 4.28$\%$ in AUC compared to the best baseline. Predicting the evolution of very sparse graphs (e.g., FB) or long-term sequences (e.g., HepPh) indeed are hard tasks. Notably, our proposed HTGN obtains remarkable gains for these datasets and successfully pushes the performance to a new level. 

It is worthwhile mentioning that all edges in the test set are new in the DISEASE dataset, which then requires the model's stronger inductive learning ability. Despite the difficulty, the proposed HTGN still outperforms the baselines by large margins and achieves notable results in both AUC and AP metrics. %\textcolor{red}{It is worthwhile mentioning that all edges in the test set are new in the DISEASE dataset, so the two link prediction tasks are equivalent. Despite the difficulty, our HTGN model achieves notable results on both AUC and AP metrics.} %Also noted that all of the links in the test snapshots of the DISEASE dataset do not appear before, which demonstrates our proposal has strong inductive learning ability. 

%We note that our proposed methods improve link prediction more substantially which shows that it can capture temporal trends better than the competing methods.
\subsubsection{New Link Prediction} New link prediction aims to predict the appearance of new links, which is more challenging. Note that static methods are not applicable to this task as the sequential order is omitted in the learning procedure of a static method, and GAE and VGAE are thus not evaluated. From Table~\ref{tab:result_new_link_prediction}, we are able to find similar observations to the link prediction task, demonstrating the superiority of the proposed HTGN. Specifically, we notice that the performance of each method drops by different degrees compared to the corresponding link prediction task, while our HTGN model produces more consistent results. For instance, the performance of the baselines degrades dramatically on AS733 (e.g., the second-best, GRUGCN drops from 94.64$\%$ to 83.14$\%$), but our HTGN only declines about 2$\%$, which shows our proposed HTGN strong inductive ability.
%\textcolor{red}{Specially, it is .... We discovery that HTGN performs much better on AS733 and the gains up to +11.40$\%$ compared with previous state-of-art method for AUC and +7.24$\%$ for AP, which is mainly from the HTC modules discussed in section~\ref{sec:ab_study}.}

%“We use micro and macro averaged AUC as evaluation metrics. Micro-AUC is calculated across the link instances from all the time steps while Macro-AUC is computed by averaging the AUC at each time step. Our results (Table 2) indicate that DySAT achieves consistent gains of 4–5$\%$ macro-AUC, in comparison to the best baseline across all datasets. Considering that the performance gain reported in other graph representation learning papers [34, 35] is usually around 2$\%$, this improvement is significant”

\subsection{Ablation Study (RQ2)}
\label{sec:ab_study}
We further conduct an ablation study to validate the effectiveness of the main components of our proposed model. We name the HTGN variants as follows:
\begin{itemize}
    \item \textbf{w/o HTC}: HTGN without the hyperbolic temporal consistency, i.e., the model is trained by minimizing the hyperbolic homophily loss only. 
    \item \textbf{w/o HTA}: HTGN without the temporal attention module, i.e., the HGRU unit directly takes the hidden state of the last timestamp as the input. 
    \item \textbf{w/o $\mathbb{H}$}: HTGN without hyperbolic geometry where all modules and learning processes are built-in Euclidean space. Correspondingly, the HTA and HTC modules are converted to Euclidean versions.
\end{itemize}
We repeat each experiment five times and report the average AUC on the test set for the link prediction task, as shown in Table~\ref{tab:ablation_study}. We first make the wrap-up observation that removing any of the components will cause performance degradation, which highlights the importance of each component. In the following, we take a closer look into the details about \textbf{w/o HTC} and \textbf{w/o HTA} at first. Discussion of \textbf{w/o $\mathbb{H}$} is present in section \ref{sec:merit_hy}.
% we go into the detail

\input{each_parts/ablation_table}
\textbf{Benefit from the HTC module.} The effect of the temporal consistency constraint is significant as the performance drops vastly if the HTC module is removed. The model degradation is significant even for long-term prediction tasks (i.e., HepPh and AS733), which confirms that the HTC module facilitates the proposed model to capture the high-level temporal smoothness of an evolving network and ensures more stable and generalized prediction performance.
 %HTC is necessary in temporal graph modeling, since the evolution of the network has local consistency. Without HTC, we found that the performance of the model has drop greatly as shown in Table ~\ref{tab:ablation_study}, especially for either low-hyperboloicity (e.g., DISEASE) datasets or long-term prediction (e.g., HepPh), which verify the effectiveness of the proposed module. 

\textbf{Benefit from the HTA module.} As observed from Table~\ref{tab:ablation_study}, the performance decays by different extents if the HTA module is removed. In particular, the degradation is the most severe on the DISEASE dataset, which assembles node attributes. It confirms our assumption that HTA is able to collect the contextual information carried in the previous snapshots to further impel the learning of HTGN.  
\section{Discussion}
In this section, we further analyze HTGN with the aim of answering the following research questions:
\begin{itemize}
    \item \textbf{RQ3}. What does hyperbolic geometry bring?
    \item \textbf{RQ4}. How is the learning efficiency in large networks?
\end{itemize}

\subsection{Merits of Hyperbolic Geometry (RQ3)}
\label{sec:merit_hy}
\textbf{Hierarchical awareness}. We remove the hyperbolic geometry and build the learning process in Euclidean space. The HTA and HTC modules are converted to the corresponding Euclidean versions. As shown in Table~\ref{tab:ablation_study} in the row for $\textbf{w/o}~\mathbb{H}$, we know that the introduction of hyperbolic geometry significantly improves the performance. Particularly, for the pure tree-like DISEASE dataset, removing the hyperbolic projection will cause the AUC to degrade about 21.35$\%$, and for another low-hyperbolicity dataset, HepPh,the deterioration is also significant with an AUC drop of about 9$\%$. It verifies that hyperbolic geometry enables the preservation of the hierarchical layout in the graph data naturally and assists in producing high-quality node representations with smaller distortion. 
%As shown in Table~\ref{tab:ablation_study}, without hyperbolic geometry, the performance of pure tree-like DISEASE dataset drops significantly and other low-hyperbolicity datasets also drop greatly which show the role of hyperbolic geometry to extract implicitly hierarchical origination. 

\begin{figure}[t]
\centering
\includegraphics[width=0.98\linewidth]{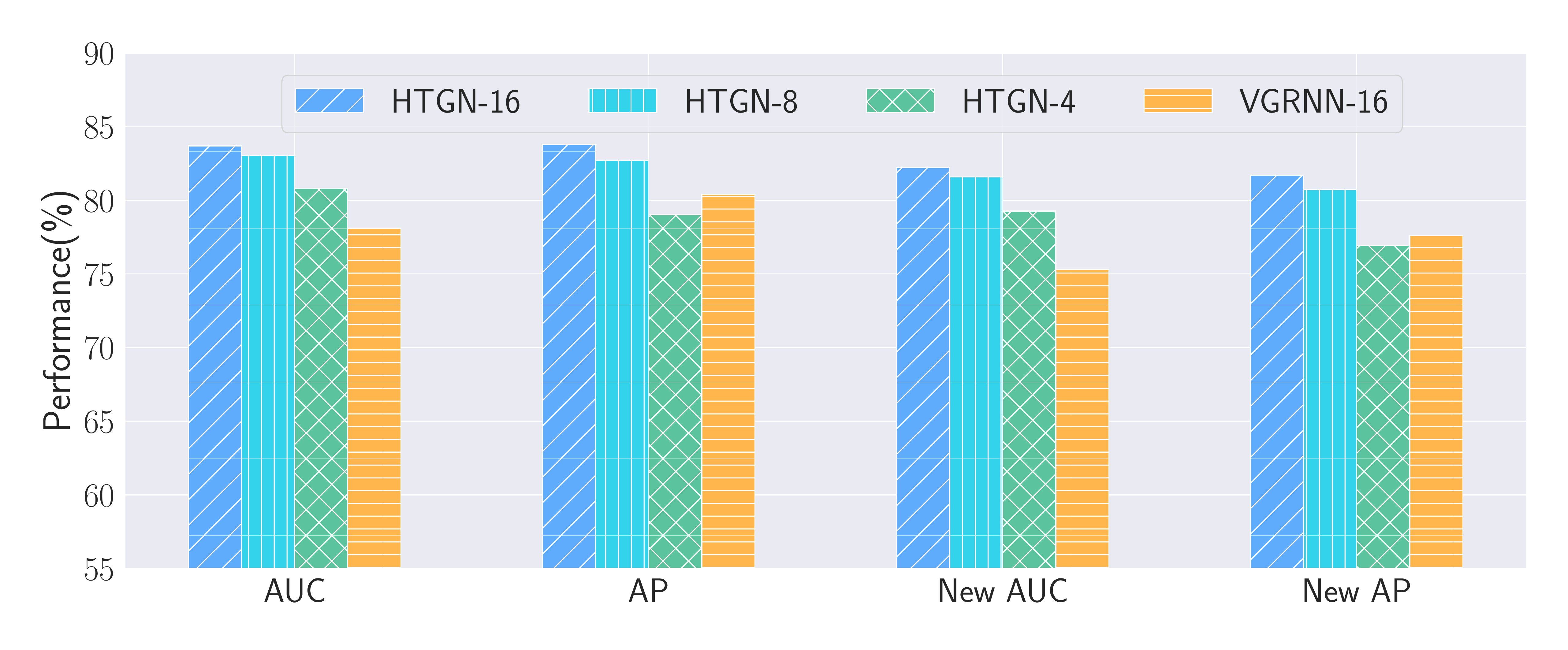}
\caption{AUC scores of different embedding dimensions on FB.}
\label{fig:embedding_dim}
\end{figure}

\textbf{Low-dimensional embedding.} Benefiting from the exponential capacity of the hyperbolic space, we are permitted to use embeddings with lower dimensions to achieve notable performance. Taking the large dataset FB as an example, as shown in Figure~\ref{fig:embedding_dim}, HTGN equipped with an 8-dimension embedding space still outperforms the runner-up, VGRNN. With 4-dimension embedding, HTGN is comparable with the 16-dimension VGRNN. This is another benefit that hyperbolic space can bring to the temporal graph network, i.e., reducing the embedding space and the corresponding learning parameters, which is valuable for embedding large-scale temporal graphs or deploying on low-memory/storage devices, e.g., mobile phones and UAV embedded units.
\begin{figure}[t]
\label{fig:ablations}
\centering
% trim={<left> <lower> <right> <upper>}
    % \includegraphics[trim={0.5cm 0cm 1.1cm 0.8cm}, clip, width=0.95\linewidth]{figures/real_time/running_time.pdf}
\includegraphics[width=0.98\linewidth]{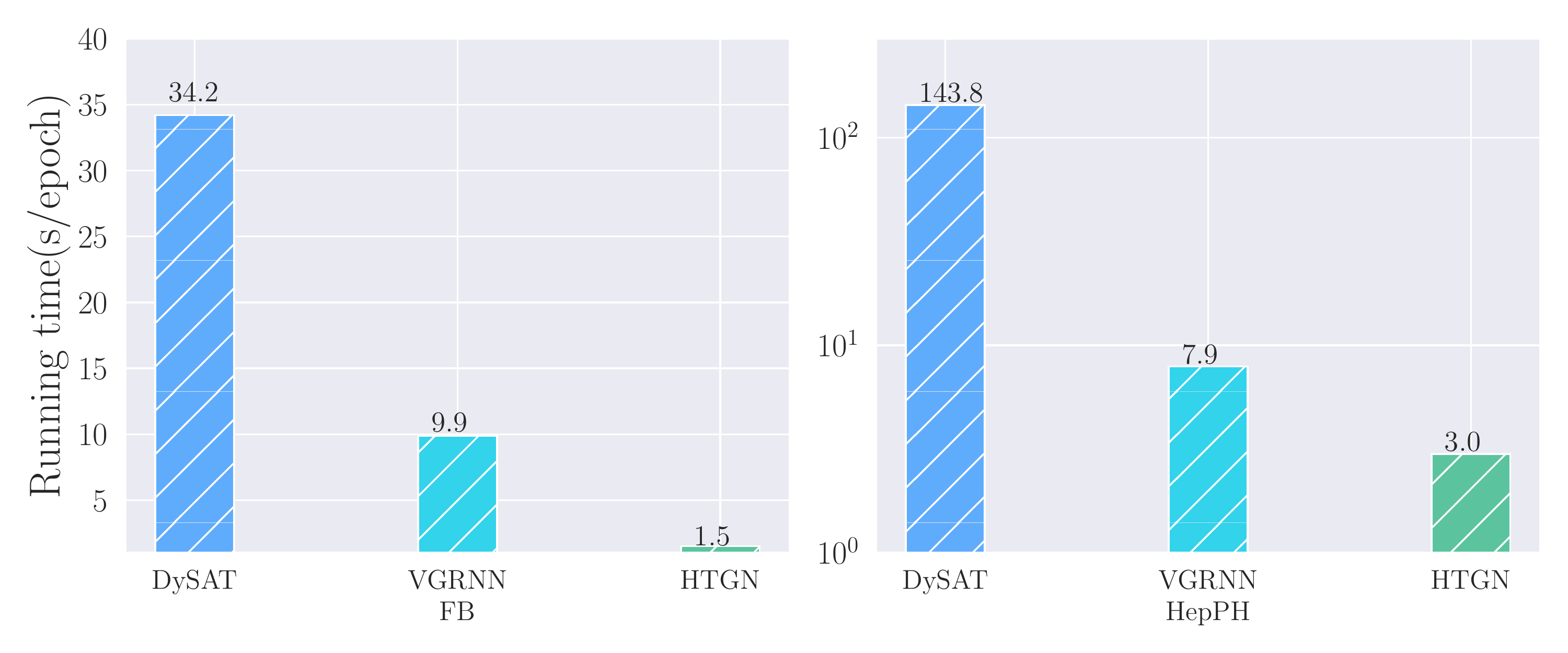}
\caption{Running time on FB and HepPh.}
\label{fig:running_time}
\vspace{-10pt}
\end{figure}

\subsection{Running Time Comparison (RQ4)}
\label{sec:scalability}
In terms of efficiency, a theoretical analysis has been presented in Section~\ref{sec:complexity}. Here, we further numerically verify the scalability of HTGN by comparing the running time with the two second-best methods: VGRNN and DySAT. VGRNN is a GRNN-based method similar to HTGN, while DySAT uses attention to capture both the spatial and temporal dynamics. Figure~\ref{fig:running_time} depicts the runtime per epoch of the three models on FB and HepPH, using a machine with GPU NVIDIA GeForce GTX TITAN X and 8 CPU cores. FB is a large-scale network with 45,435 nodes, and HepPH has fewer nodes 15,330 but more connection links.

As observed, HTGN achieves substantially lower training times, e.g., the running time per epoch of HTGN is 1.5 seconds compared to 9.9s for VGRNN and 34.2s for DySAT. The main reason is that HTGN deploys a shared HGNN before feeding into HGRU while VGRNN utilizes different GNNs\footnote{https://github.com/VGraphRNN/VGRNN/blob/master/VGRNN\_prediction.py}. On the other hand, DySAT requires computing both temporal attention and structural attention, which is computationally heavy. For the more dense HepPH network, the computing cost of DySAT which utilizes a pure self-attentional architecture increases dramatically, demonstrating the efficient recurrent learning paradigm in dynamic graph embedding.
%To further intuitively compare the running efficiency of HTGN and other temporal graph models, we list the running time in Table~\ref{fig:running_time} of the two largest datasets, i.e., HepPH and FB where the results are averaged 10 runs for total 100 epochs. From the real-running time, we easily know that HTGN is quite efficient. The main reason is that we use the same HTGN cells with sharing parameters at all timestamps, and deploy an HGNN in the HTGN cell while VGRNN and CGRN utilize four different GCNs.

\section{Conclusion}
%Research on the temporal network representation has much significance since it helps us to understand the system behaviours and predict further activities. Although, a number of competing models were proposed to model the temporal networks, the complexity and hierarchical nature was underestimated. To make up for this gap, we proposed a novel model by embedding the network structure into the hyperbolic space due to its large capacity and hierarchical awareness. Further, we combine it with graph neural network, recurrent neural network to implicitly learn the evolving patterns of temporal networks. The experimental and ablation study demonstrate that hyperbolic geometry and our proposed hyperbolic modules play a very crucial role, which provides a new direction for the study of temporal graph networks and is of great significance. Treating a temporal network as a series snapshots is a discrete-time learning branch modeling temporal network, in the future we will focus on continuous representation learning to explore more possibilities. 
% Hyperbolic embeddings excel in their encoding of both node similarity and hierarchy, which allows for unsupervised identification to extract latent hierarchical structures. 
In this work, we introduce a novel hyperbolic geometry-based node representation learning framework, denoted as hyperbolic temporal graph network, HTGN, for temporal network modeling. In general, HTGN follows the concise and effective GRNN framework but leverages the power of hyperbolic graph neural network and facilitates hierarchical arrangement to capture the topological dependency. More specifically, two novel modules: hyperbolic temporal contextual self-attention (HTA) and hyperbolic temporal consistency (HTC), respectively extract attentive historical states and ensuring stability and generalization, are proposed to impel the success of HTGN. 
% To the best of our knowledge, this is the first work to address the temporal graph embedding via a  recurrent learning scheme built-in hyperbolic space. 
When evaluated on multiple real-world temporal graphs, our approach outperforms the state-of-the-art temporal graph embedding baselines by a large margin. For future work, we will generalize our method to more challenging tasks and explore continuous-time learning to incorporate the fine-grained temporal variations.
%\clearpage

\section*{Acknowledgements}
This work is partially supported by the National Key Research and Development Program of China (No. 2018AAA0100204) and CUHK 3133238, Research Sustainability of Major RGC Funding Schemes (RSFS). We would like to thank the anonymous reviewers for their constructive comments.

\bibliographystyle{ACM-Reference-Format}
\bibliography{reference}
\clearpage
%\section{Appendix}
\input{each_parts/appendix}

\end{document}

%% file: each_parts/aggregation.tex
Similar to~\cite{hgcn2019,liu2019HGNN,zhang2019hyperbolic}, we perform the aggregation computation in the tangent space. We adopt the attention-based aggregation, which is formulated as:
\begin{equation}
\small
\begin{aligned}
\alpha_{ij}&=softmax_{(j\in\mathcal{N}(i))}(s_{ij})=\frac{\exp(s_{ij})}{\sum_{j'\in\mathcal{N}_i} \exp(s_{ij'})}, \\
s_{ij}&=\mathrm{LeakReLU}(a^T[\log_{0}^c(m_i^{l})\|\log_{0}^c(m_j^{l})]),
\end{aligned}
\end{equation}
where $a$ is a trainable vector, $\|$ denotes concatenation operation and $s_{ij}$ indicates the correlation between the neighbors $j\in\mathcal{N}(i)$ and the center node $i$. Besides, we also consider the Laplacian based method~\cite{hgcn2019}, that is $\alpha_{ij}=\frac{1}{\sqrt{(d_i+1)(d_j+1)}}$ where $d_i$ and $d_j$ are the degree of node $i$ and node $j$.

\begin{comment}

Similar to~\cite{hgcn2019}, we perform the aggregation computation in the tangent space, considering the following two aggregation approaches: %Laplacian-based~\cite{gcn2017} and attention-based aggregation~\cite{GAT}: %Discussion and equations of the two aggregation operations are presented in Appendix~\ref{appendix:agg}.
%\begin{comment}
%, defined as follows:    \\
\noindent
(1) Laplacian-based aggregation~\cite{gcn2017}:
\begin{equation}
    \alpha_{ij} = \frac{1}{\sqrt{d_id_j}},
\end{equation}
where $d_i=|\mathcal{N}_i|$ denotes the degree of node $i$.\\
(2) Attention-based aggregation~\cite{GAT}:
\begin{equation}
\begin{aligned}
\alpha_{ij}&=softmax_{(j)}(s_{ij})=\frac{\exp(s_{ij})}{\sum_{j'\in\mathcal{N}_i} \exp(s_{ij'})}, \\
s_{ij}&=\mathrm{LeakReLU}(a^T[\log_{0}^c(m_i^{l})\|\log_{0}^c(m_j^{l})]),
\end{aligned}
\end{equation}
where $a$ is a trainable vector, $\|$ denotes concatenation operation and then $e_{ij}$ indicates the importance of neighbors $j\in\mathcal{N}(i)$ with respect to the center node $i$.

As observed, the Laplacian-based aggregation only relies on the network structure but the attention-based is more associated with the node features. For datasets with informative node attributes, the attention-based approach generally achieves promising results~\cite{GAT,sankar2020dysat}.\\ 
\end{comment}

%% file: each_parts/proposition.tex
\noindent
%\textbf{Remarks}. 
\begin{prop}
\label{prop:loss}
Let $N$ be the number of nodes, $T$ be the number of training timestamps, $\mathcal{N}(i)$ be the neighbors of node $i$, and $|\mathcal{E}_t|$ be the number of links in timestamp $t$. Then, minimizing the loss $\mathcal{L}=\sum_{t=1}^T\mathcal{L}_t$
is equivalent to (1) minimizing the hyperbolic distance of a node with its current and historically connected nodes, and maximizing with the sampled negative neighbors, which are weighted by $\frac{1}{|\mathcal{E}_t|}$; (2) minimizing the distance between the same node over two consecutive timestamps, that is:\\
\begin{equation}
\small
\begin{aligned}
\mathcal{L}&
    =\sum_{i}^N\sum_{t=1}^T\left(\sum_{j\in\mathcal{N}(i)} \frac{1}{|\mathcal{E}_t|} d^\mathcal{H}(\mathbf{x}_{t,i}^\mathcal{H}, \mathbf{x}_{t,j}^\mathcal{H})-\sum_{j\notin\mathcal{N}(i)} \frac{1}{|\mathcal{E}_t|}d^\mathcal{H}(\mathbf{x}_{t,i}^\mathcal{H}, \mathbf{x}_{t,j'}^\mathcal{H})\right.\\
    &\quad\quad\quad\quad\quad\quad\quad\quad\quad\quad\quad\quad\quad\quad\quad\left.+\frac{\lambda}{N}d^\mathcal{H}(\mathbf{x}_{t,i}^\mathcal{H}, \mathbf{x}_{t-1, i}^\mathcal{H})\right).
\end{aligned}
\end{equation}
\end{prop}
\noindent
\textit{Proof.} See Appendix~\ref{appendix:prop2}. 

Proposition 2 shows that our loss builds a message-passing connection within its neighbors from different times and local structures. In other words, if two nodes are not connected directly, they may still have implicit interactions through their common neighbors even at different times. This enables the representation to encode more patterns directly and indirectly, which is essential for high-quality representation and further prediction. 

As we can see, the loss function $\mathcal{L}_t$ is only related to distance in the Poincar{\'e} ball, and thus scales well to large-scale datasets. Given two points $u, v\in\mathbb{H}^{d'}$, the gradient~\cite{nickel2017poincare} of their distance in the Poincar{\'e} Ball is formulated as:
\begin{equation}
\small
    \Delta_u(d(u,v))=\frac{4}{\beta\sqrt{\gamma^2-1}}\left(
    \frac{\|u\|^2-2<u,v>+1}{\alpha^2}u-\frac{v}{\alpha}
    \right),
\end{equation}
where $\alpha=1-\|u\|^2, \beta=1-\|v\|^2, \gamma=1+\frac{2}{\alpha\beta}\|u-v\|^2$. 
The computational and memory complexity of one backpropagation step depends linearly on the embedding dimension.

%% file: each_parts/auc_ap_table.tex
\begin{table*}[t]
\centering
\caption{AUC (left) and AP (right) scores of temporal link prediction on temporal network datasets.}
\label{tab:result_link_prediction}
\resizebox{\textwidth}{!}{%
\begin{tabular}{@{}lcccccccccccc@{}}

\toprule
\multicolumn{7}{c}{\textbf{AUC}}    & \multicolumn{6}{c}{\textbf{AP}}       \\
\textbf{Dataset}    & \textbf{DISEASE}    & \textbf{HepPh}      & \textbf{FB}         & \textbf{AS733}      & \textbf{Enron}      & \multicolumn{1}{c|}{\textbf{COLAB}}       & \textbf{DISEASE}    & \textbf{HepPh}      & \textbf{FB}         & \textbf{AS733}      & \textbf{Enron}      & \textbf{COLAB}       \\ \midrule \midrule
\textbf{GAE}        & $72.55\pm1.20$          & $69.44\pm0.56$          & $63.07\pm0.93$          & $93.21\pm1.53$          & $92.50\pm0.68$          & \multicolumn{1}{c|}{$84.57\pm0.64$}           & $60.55\pm1.01$          & $73.61\pm0.58$          & $65.35\pm0.90$          & $94.75\pm0.90$          & $\mathit{93.48\pm0.64}$          & $87.69\pm0.44$          \\
\textbf{VGAE}       & $83.08\pm1.27$          & $72.39\pm0.11$          & $67.16\pm0.53$          & $\mathit{95.76\pm0.91}$ & $91.93\pm0.34$          & \multicolumn{1}{c|}{$85.16\pm0.74$}           & $78.34\pm1.25$          & $75.78\pm0.06$          & $69.73\pm0.17$          & $96.42\pm0.55$          & $93.45\pm0.49$          & $88.70\pm0.35$          \\
\textbf{EvolveGCN}  & $73.55\pm4.23$          & $76.82\pm1.46$          & $76.85\pm0.85$          & $92.47\pm0.04$          & $90.12\pm0.69$          & \multicolumn{1}{c|}{$83.88\pm0.53$}           & $73.25\pm3.44$          & $81.18\pm0.89$          & $80.87\pm0.64$          & $95.28\pm0.01$          & $92.71\pm0.34$          & $87.53\pm0.22$          \\
\textbf{GRUGCN}     & $79.25\pm1.69$          & $\mathit{82.86\pm0.53}$ & $\mathit{79.38\pm1.02}$ & $94.96\pm0.35$          & $92.47\pm0.36$          & \multicolumn{1}{c|}{$84.60\pm0.92$}           & $65.26\pm1.94$          & $\mathit{85.87\pm0.23}$ & $\mathit{82.77\pm0.75}$ & $96.64\pm0.22$          & $93.38\pm0.24$          & $87.87\pm0.58$          \\
\textbf{DySAT}      & $73.74\pm2.28$          & $81.02\pm0.25$          & $76.88\pm0.08$          & $95.06\pm0.21$          & $93.06\pm0.97$ & \multicolumn{1}{c|}{$\mathit{87.25\pm1.70}$}  & $63.81\pm1.86$          & $84.47\pm0.23$          & $80.39\pm0.14$          & $\mathit{96.72\pm0.12}$ & $93.06\pm1.05$          & $\mathit{90.40\pm1.47}$          \\ 
\textbf{VGRNN}      & $\mathit{86.44\pm3.12}$ & $77.65\pm0.99$          & $78.11\pm1.11$          & $95.17\pm0.62$          & $\mathit{93.10\pm0.57}$          & \multicolumn{1}{c|}{$85.95\pm0.49$}           & $\mathit{82.00\pm3.83}$ & $80.95\pm0.94$          & $80.40\pm0.74$          & $96.69\pm0.31$          & $93.29\pm0.69$ & {$87.77\pm0.79$} \\ \midrule
\textbf{HTGN (Ours)}& $\mathbf{89.65\pm0.70}$ & $\mathbf{91.13\pm0.14}$ & $\mathbf{83.70\pm0.33}$ & $\mathbf{98.75\pm0.03}$ & $\mathbf{94.17\pm0.17}$ & \multicolumn{1}{c|}{$\mathbf{89.26\pm0.17}$}  & $\mathbf{84.63\pm0.65}$ & $\mathbf{89.52\pm0.28}$ & $\mathbf{83.80\pm0.43}$ & $\mathbf{98.41\pm0.03}$ & $\mathbf{94.31\pm0.26}$ & $\mathbf{91.91\pm0.07}$ \\ 
\textbf{Gain (\%)}  & +3.71               & +9.98               & +5.44               & +3.12               & +1.15               & \multicolumn{1}{c|}{+2.30}               & +3.21                      & +4.25               & +1.24               & +1.75               & +0.89             & +1.67               \\ \bottomrule
\end{tabular}% 
}
\end{table*}

\begin{table*}[t]
\centering
\caption{AUC (left) and AP (right) scores of temporal new link on temporal network datasets.}
\label{tab:result_new_link_prediction}
\resizebox{\textwidth}{!}{%
\begin{threeparttable}
\begin{tabular}{@{}lcccccccccccc@{}}

\toprule
\multicolumn{7}{c}{\textbf{AUC}}                                                                                                                                            & \multicolumn{6}{c}{\textbf{AP}}                                                                                                  \\
\textbf{Dataset}   & \textbf{DISEASE}$^\star$    & \textbf{HepPh}       & \textbf{FB}         & \textbf{AS733}      & \textbf{Enron}      & \multicolumn{1}{c|}{\textbf{COLAB}}       & \textbf{DISEASE}   & \textbf{HepPh}      & \textbf{FB}          & \textbf{AS733}      & \textbf{Enron}      & \textbf{COLAB}       \\ \midrule \midrule
\textbf{EvolveGCN} & $73.55\pm4.23$          & $74.79\pm1.61$           & $74.49\pm0.89$          & $75.82\pm0.67$          & $82.85\pm0.97$          & \multicolumn{1}{c|}{$73.49\pm0.86$}          & $73.25\pm3.44$         & $79.04\pm1.02$          & $78.33\pm0.66$           & $83.57\pm0.46$          & $85.01\pm0.22$          & $77.11\pm0.44$          \\
\textbf{GRUGCN}    & $79.25\pm1.69$          & $\mathit{81.97\pm0.49}$  & $\mathit{77.69\pm1.03}$ & $\mathit{83.14\pm1.21}$          & $87.59\pm0.57$          & \multicolumn{1}{c|}{$75.60\pm1.60$}          & $65.26\pm1.94$         & $\mathit{84.78\pm0.22}$ & $\mathit{81.07\pm0.77}$& $88.14\pm0.76$          & $\mathit{88.41\pm0.45}$          & $78.55\pm1.05$          \\
\textbf{DySAT}     & $73.74\pm2.28$          & $79.01\pm0.26$           & $74.97\pm0.12$          & $82.84\pm0.72$          & $87.94\pm3.78$ & \multicolumn{1}{c|}{$\mathit{79.74\pm4.35}$} & $63.81\pm1.86$                 & $82.53\pm0.25$          & $78.34\pm0.07$         & $\mathit{89.07\pm0.57}$ & $86.83\pm5.01$          & $\mathit{83.47\pm3.01}$ \\ 
\textbf{VGRNN}     & $\mathit{86.44\pm3.12}$ & $74.52\pm1.05$         & $75.31\pm1.10$          & $78.86\pm3.39$           & $\mathit{88.43\pm0.75}$          & \multicolumn{1}{c|}{$77.09\pm0.23$}          & $\mathit{82.00\pm3.83}$ & $77.83\pm0.93$          & $77.61\pm0.64$         & $84.59\pm1.90$          & $87.57\pm0.57$  & $79.63\pm0.94$ \\ \midrule
\textbf{HTGN (Ours)}& $\mathbf{89.65\pm0.70}$ & $\mathbf{90.11\pm0.14}$ & $\mathbf{82.21\pm0.41}$ & $\mathbf{96.62\pm0.22}$ & $\mathbf{91.26\pm0.27}$ & \multicolumn{1}{c|}{$\mathbf{81.74\pm0.56}$} & $\mathbf{84.63\pm0.65}$ & $\mathbf{88.18\pm0.31}$ & $\mathbf{81.70\pm0.46}$ & $\mathbf{95.52\pm0.25}$ & $\mathbf{90.62\pm0.34}$ & $\mathbf{84.06\pm0.41}$ \\ 
\textbf{Gain (\%)} & +3.71               & +9.93                & +5.82                & +11.40               & +3.20                & \multicolumn{1}{c|}{+2.51}                & +3.21                & +4.01                & +0.78              & +7.24               & +2.5                & +0.71                \\ \bottomrule
\end{tabular}%
\begin{tablenotes}
        %\footnotesize
        \small
        \item[$\star$] In the DISEASE dataset, all edges in the test set are new, so the results of new link prediction are equivalent to the results of link prediction.
    \end{tablenotes}
\end{threeparttable}
}
\end{table*}

%% file: each_parts/ablation_table.tex
% \usepackage{booktabs}
% \usepackage{multirow}
% \usepackage{graphicx}
\begin{table}[t]
\centering
\caption{Ablation study.}
\label{tab:ablation_study}
\resizebox{0.475\textwidth}{!}{%
\begin{tabular}{@{}llcccccc@{}}
\toprule
\multicolumn{2}{c}{\textbf{Dataset}}        & \textbf{DISEASE}    & \textbf{HepPh}      & \textbf{FB}         & \textbf{AS733}      & \textbf{Enron}      & \textbf{COLAB}       \\ 
\multicolumn{2}{c}{\textbf{$\delta$}}       & 0.0                 & 1.0                 & 2.0                 & 1.5                 & 1.5                 & 2.0                 \\
\multicolumn{2}{c}{\textbf{$\rho(0.01)$}}   & 0.41                & 1.37                & 0.04                & 0.2                 & 3.37                & 0.94                \\
                % & \textbf{91.76+0.40} & \textbf{91.13+0.14} & \textbf{83.70+0.33} & \textbf{98.75+0.03} & \textbf{93.92+0.07} & \textbf{89.26+0.17} 
 \midrule \midrule
& \textbf{w/o HTC}          & $78.22\pm0.26$    & $76.30\pm0.95$    & $79.74\pm0.34$    & $95.07\pm0.63$    & $92.88\pm0.24$    & $85.16\pm0.28$          \\
& \textbf{Gain (\%)}        & -14.6        & -19.44        & -4.97         & -3.87         & -1.39         & -4.81                \\ \cmidrule(l){2-8}
& \textbf{w/o HTA}          & $83.51\pm1.71$    & $90.87\pm0.10$    & $82.97\pm0.44$   & $98.00\pm0.05$    & $93.51\pm0.19$    & $88.60\pm0.20$          \\
& \textbf{Gain (\%)}        & -7.35        & -0.29         & -0.88         & -0.77         & -0.71         & -0.74                \\ \cmidrule(l){2-8}
& \textbf{w/o} $\mathbb{H}$   & $73.88\pm2.66$	& $83.33\pm0.47$	& $80.53\pm0.24$	& $95.12\pm0.36$	& $92.28\pm0.32$	& $84.20\pm0.79$          \\
& \textbf{Gain (\%)}        &-21.35	        &-9.36	        &-3.94	        &-3.82	        &-2.05	        &-6.01                \\ 
                         \bottomrule
% \multirow{6}{*}{\textbf{LP_{new}}} & \textbf{w/o HTC}   & 78.22+0.26     & 74.74+0.92    & 77.92+0.38    & 87.70+0.43    & 87.67+0.20    & 77.17+0.45          \\
%                             & \textbf{Gain (\%)}        & -17.31        & -20.40        & -4.85         & -8.81         & -2.76         & -8.46               \\ \cmidrule(l){2-8}
%                             & \textbf{w/o HTA}          & 83.51+1.71    & 89.85+0.11    & 81.46+0.57    & 94.91+0.19    & 89.47+0.40    & 81.19+0.52          \\
%                             & \textbf{Gain (\%)}        & -9.88         & -0.16         & -0.29         & -0.64         & -0.69         & -3.09                \\ \cmidrule(l){2-8} 
%                             & \textbf{w/o} $\mathbb{H}$ & 73.88+2.66    & 82.43+0.41    & 78.87+0.24    & 83.43+0.83    & 86.91+1.01    & 76.92+2.44          \\
%                             & \textbf{Gain (\%)}        & -24.20         & -9.17         & -3.59         & -14.49        & -3.66         & -8.81                \\ 
%                          \bottomrule
\end{tabular}%
}
\vspace{-10pt}
\end{table}

%% file: each_parts/appendix.tex
\appendix
\section{Appendix}

% \begin{comment}
\subsection{Geometry Initiations of Hyperbolic Space}
Riemannian manifolds with different curvatures define different geometries, where curvature measures how much a geometric object deviates from a flat plane, or in the case of a curve, deviates from a straight line. Different from the well-known Euclidean geometry which has zero curvature, hyperbolic space is a type of manifold with constant negative curvature and thus shows distinguishing properties.

There exist multiple equivalent models for hyperbolic space. The most commonly used in the machine learning community are the Poincaré (disk) model and the Lorentz (hyperboloid) model. The Lorentz model is well-suited for Riemannian optimization and the Poincaré disk provides a very intuitive way for visualizing and interpreting hyperbolic embeddings. We here take Poincaré disk to illustrate some intuitions of hyperbolic space. %, whose distance metric is functioned by Equation(~\ref{equ:poincare}). 

\begin{figure}[htpb]
\centering
\includegraphics[width=0.98\linewidth]{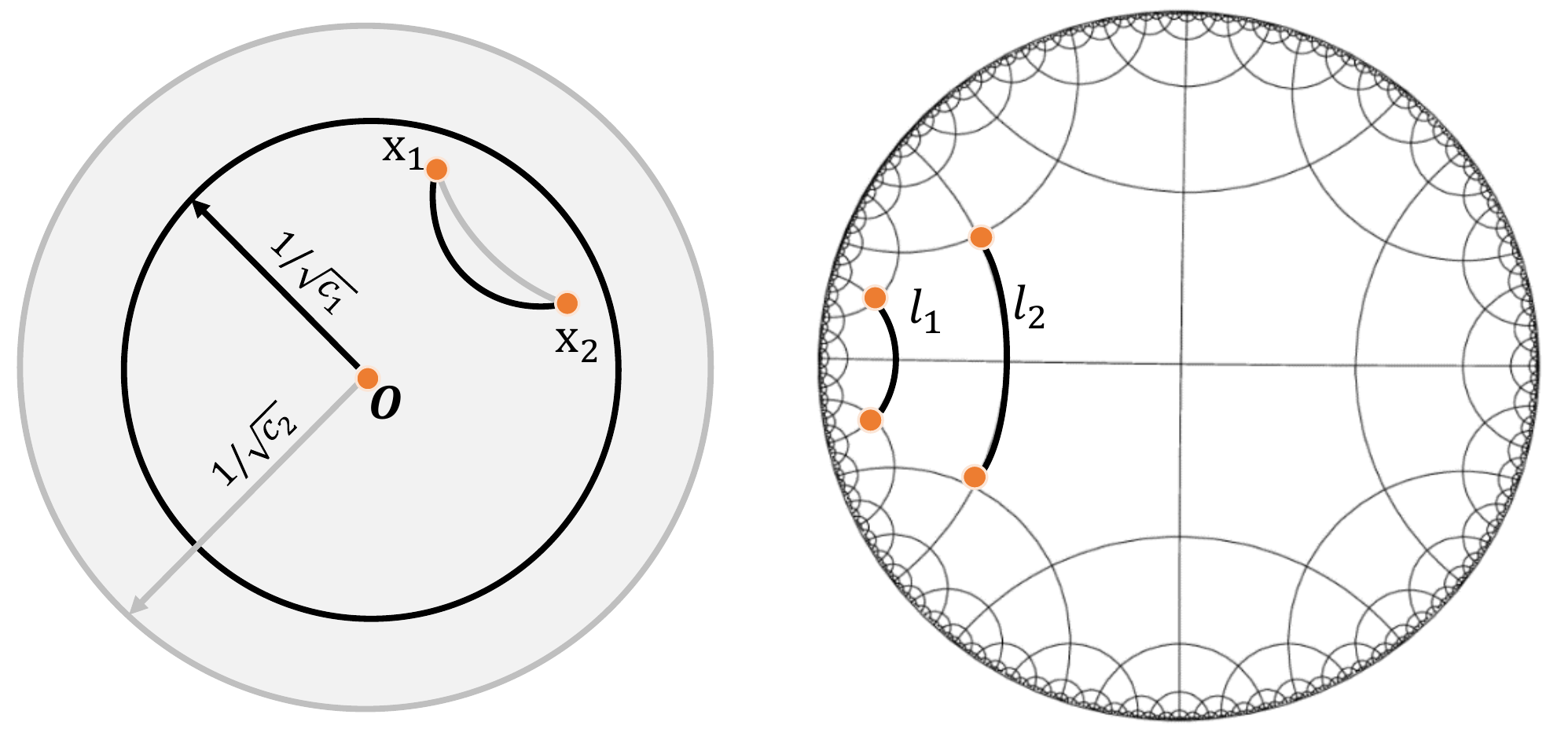}
\caption{\textbf{Left:} Visualizations of Poincar\'e disks with different curvatures $-c_1 (c_1>0)$ (area within the black circle and the corresponding radius is $1/\sqrt{c_1})$), $-c_2$ (area within the grey circle and the corresponding radius is $1/\sqrt{c_2}(c_2>0)$), where $\mathbf{x}_1$ and $\mathbf{x}_2$ are two points on Poincar\'e disks. Right: Two lines $l_1$ and $l_2$ with the same length in Poincar\'e disk.}
\label{fig:poincare}
\end{figure}

Figure~\ref{fig:poincare} gives some visualizations of the geometry properties on the Poincar{\'e} disk.  %shows the geodesics(the shortest path) of two points on  Poincar{\'e} disk with different curvatures(left) and the hyperbolic lines with same hyperbolic length(right).
As observed on the left, the geodesic length of two points in different Poincar\'e disks is different and related to its curvature. When the radius $1/\sqrt{c}$ decreases (i.e., space bends more or the absolute value of curvature increases), the distance between two given points will increase, and the line is closer to the origin. From the right, $l_1$ and $l_2$ are two lines with the same length though, $l_1$ is shorter from our Euclidean view. Then when two lines are with the same ``Euclidean'' length, the one closer to the border is actually longer in hyperbolic space and can accommodate more. We further give some mathematical expressions to illustrate the exponentially increased capacity of hyperbolic space.

According to \cite{lozier2003nist}, the $n$-dimensional volume of a Euclidean ball of radius $r$ is:
\begin{equation}
\small
   V^{\mathbb{E}}_{n}(r)=\frac{\pi^{\frac{n}{2}}}{\Gamma\left(\frac{n}{2}+1\right)} r^{n}.
    \label{equ:vol_euclidean}
\end{equation}
While the $n$-dimensional volume of a hyperbolic space of radius $r$ in $n$-dimensional hyperbolic space, referring to~\cite{bridson2013metric}, is given as: % (denoted as $V^{\mathbb{E}}_{n}(r)$) and the $n$-dimensional volume of a Poincaré ball of radius $r$ (denoted as $V^{\mathbb{B}}_{n}(r)$) are given by~\cite{bridson2013metric}:
\begin{equation}
\small
V^{\mathbb{H}}_{n}(r) =V\left(\mathbb{S}^{n-1}\right) \int_{0}^{r} \sinh ^{n-1} t d t,
\end{equation}
where $V\left(\mathbb{S}^{n-1}\right)$ is the volume of the tangent space centered on the origin of the Poincaré model of radius $r$. 
% \begin{comment}
% \begin{equation}
% \begin{aligned} 
% V^{\mathbb{H}}_{n}(r) &=\operatorname{Vol}\left(\mathbb{S}^{n-1}\right) \int_{0}^{\tanh \frac{1}{2}} \frac{2^{n} s^{n-1}}{\left(1+s^{2}\right)^{n}} d s \\ &=2^{n-1} \operatorname{Vol}\left(\mathbb{S}^{n-1}\right) \int_{0}^{r} \sinh ^{n-1} \frac{t}{2} \cosh ^{n-1} \frac{t}{2} d t \\ &=\operatorname{Vol}\left(\mathbb{S}^{n-1}\right) \int_{0}^{r} \sinh ^{n-1} t d t 
% \end{aligned}
% \end{equation}
% \end{comment}
In the 2-dimensional case, we then have the explicit expression  $V^{\mathbb{H}}_{n}(r)=4\pi \sinh^2(\frac{r}{2})$. For all $x \in \mathbb{H}^n$, we have
\begin{equation}
   V^{\mathbb{H}}_{n}(r) \sim \frac{\operatorname{Vol}\left(\mathbb{S}^{n}\right)}{2^{n-1}} e^{(n-1) r}, as\quad r\rightarrow\infty.
 \label{equ:vol_hyperbolic}
\end{equation}

%as $r\rightarrow\infty$.

As concluded from equations \eqref{equ:vol_euclidean} and \eqref{equ:vol_hyperbolic}, the volume of a ball in hyperbolic space grows exponentially with the radius, while the counterpart in Euclidean spaces expands polynomially. Meanwhile, the nodes of a tree also grows exponentially with the depth (e.g., a perfect binary tree with depth $n$ has $2^{n+1}-1$ nodes). A hyperbolic space can thus be regarded as a continuous analogous of trees and can be applied to naturally model data with hierarchical structures or tree-like layout.  %much faster than in Euclidean space.  They expand faster than Euclidean spaces, because Euclidean spaces expand polynomially while hyperbolic spaces expand exponentially. 

%\textcolor{red}{TBA zhoumin}

%http://users.jyu.fi/~parkkone/RG2012/HypGeom.pdf
%\begin{comment}
% Figure \ref{fig:poincare} shows the examples of a poncare \' disk where left is 

% As can be seen, around the origin of the disk, these lines look straight, as in Euclidean space. However, the closer they get to the border, the stronger they bend.

% Right shows the Hyperbolic lines $l_1$ and $l_2$ in the Poincaré disk with same hyperbolic length.

% Note that these lines are geodesics, i.e. they are shortest paths between the two points that they connect, and their length in hyperbolic space gives the hyperbolic distance between the two points. All edges of this graph have same hyperbolic length. The space looks more tree-like and has more capacity at the border.

% This exponential volume growth of hyperbolic spaces is reminiscent of trees. Indeed, a binary tree of depth $n$ contains $2^n$ nodes.

% Similarly, one can think of a hyperbolic space as stretching the metric in opposite directions, connecting together the concepts of exponential volume growth, continuous tree-likeness and negative curvature. 

% The power applies to any kind of data with an approximate underlying hierarchical structure

% http://hyperbolicdeeplearning.com/simple-geometry-initiation/
%\end{comment}

%\input{each_parts/poincare_disk}
% \end{comment}

%\subsection{Details on loss function}
\subsection{Proof of Proposition~\ref{prop:loss}}
\label{appendix:prop2}
%We here provide the proof of Proposition~\ref{prop:loss} in Section~\ref{section:loss}. 
\begin{proof}
First, recall the equation of loss function $\mathcal{L}_t$:
\begin{equation}
\small
\begin{aligned}
\mathcal{L}_t & =\mathcal{L}_{t,r} + \lambda\mathcal{L}_{t,c} \\
             &  = \frac{1}{E_1}\sum_{e_{ij}\in \mathcal{E}}-\log(p_f(\mathbf{x}_i^\mathcal{H}, \mathbf{x}_j^\mathcal{H}))- \frac{1}{E_2}\sum_{e_{i'j'}\notin\mathcal{E}}(1-\log(p_f(\mathbf{x}_{i'}^\mathcal{H}, \mathbf{x}_{j'}^\mathcal{H}))) \\
            & \quad\quad\quad\quad\quad\quad\quad\quad\quad\quad\quad\quad\quad+ \lambda\cdot\frac{1}{N}\sum_{i=1}^N d^\mathcal{H}(\mathbf{x}_{t,i}^\mathcal{H}, \mathbf{x}_{(t-1),i}^\mathcal{H}). 
\end{aligned}
\end{equation}
The first term can be arranged as:
\begin{equation}
\small
\begin{aligned}
       &\frac{1}{E_1}\sum_{e_{ij}\in \mathcal{E}}-\log(p_f(\mathbf{x}_i^\mathcal{H}, \mathbf{x}_j^\mathcal{H}))\\
       &=\frac{1}{E_1}\sum_{e_{ij}\in \mathcal{E}}-\log([\exp((d^\mathcal{H}(\mathbf{x}_i^\mathcal{H}, \mathbf{x}_j^\mathcal{H})-r)/s)]^{-1}) \\
        &=\frac{1}{E_1}\sum_{e_{ij}\in \mathcal{E}}d^\mathcal{H}(\mathbf{x}_i^\mathcal{H}, \mathbf{x}_j^\mathcal{H})-r)/s.
\end{aligned}
\end{equation}
Similarly, the second term can be arranged as:
\begin{equation}
\small
\begin{aligned}
        &\frac{1}{E_2}\sum_{e_{i'j'}\notin \mathcal{E}}1-\log(p_f(\mathbf{x}_{i'}^\mathcal{H}, \mathbf{x}_{j'}^\mathcal{H})) \\
        &=\frac{1}{E_2}\sum_{e_{i'j'}\notin \mathcal{E}}1-\log([\exp((d^\mathcal{H}(\mathbf{x}_{i'}^\mathcal{H}, \mathbf{x}_{j'}^\mathcal{H})-r)/s)]^{-1})\\
        &=\frac{1}{E_2}\sum_{e_{i'j'}\notin \mathcal{E}}1+(d^\mathcal{H}(\mathbf{x}_{i'}^\mathcal{H}, \mathbf{x}_{j'}^\mathcal{H})-r)/s).
\end{aligned}
\end{equation}
Then, we have:
\begin{equation}
\small
\begin{aligned}
    \mathcal{L}_t  &=  \epsilon_0
    +\epsilon_1\sum_{i}^n d^\mathcal{H}(\mathbf{x}_{(t, i)}^\mathcal{H}, \mathbf{x}_{(t-1, i)}^\mathcal{H})+\epsilon_2\sum_{e_{ij}\in \mathcal{E}}d^\mathcal{H}(\mathbf{x}_i^\mathcal{H}, \mathbf{x}_j^\mathcal{H}) \\
    &\quad\quad\quad\quad\quad\quad\quad\quad\quad\quad\quad\quad\quad\quad-\epsilon_3\sum_{e_{i'j'}\notin \mathcal{E}}d^\mathcal{H}(\mathbf{x}_{i'}^\mathcal{H}, \mathbf{x}_{j'}^\mathcal{H}),
\end{aligned}
\end{equation}
where $\epsilon_0 = -1,\epsilon_1=\frac{\lambda}{|V|}, \epsilon_2=\epsilon_3=\frac{1}{s|\mathcal{E}_t|}$. Note that we sample the same number of negative edges as there are positive ones, hence, $E_1=E_2=|\mathcal{E}_t|$. In our experiments, $s$ is set to $1.0$. Adding up the loss of all timestamps, we have:
\begin{equation}
\small
\begin{aligned}
    \mathcal{L}&=\sum_{t=1}^T \mathcal{L}_t\\
    &=-T +\sum_{t=1}^T\sum_{i}^N \frac{\lambda}{|V|}d^\mathcal{H}(\mathbf{x}_{(t, i)}^\mathcal{H}, \mathbf{x}_{(t-1, i)}^\mathcal{H})+\sum_{t=1}^T\sum_{e_{ij}\in \mathcal{E}}\frac{1}{|\mathcal{E}_t|}d^\mathcal{H}(\mathbf{x}_i^\mathcal{H}, \mathbf{x}_j^\mathcal{H})\\ &\quad\quad\quad\quad\quad\quad\quad\quad\quad\quad\quad\quad\quad\quad\quad\quad-\sum_{t=1}^T\sum_{e_{i'j'}\notin \mathcal{E}}\frac{1}{|\mathcal{E}_t|}d^\mathcal{H}(\mathbf{x}_i^\mathcal{H}, \mathbf{x}_j^\mathcal{H})\\
    % &=\sum_{t=1}^T\sum_{i}^N\sum_{j\in\mathcal{N}_i} w_{t,j} d^\mathcal{H}(\mathbf{x}_i^\mathcal{H}, \mathbf{x}_j^\mathcal{H})-\sum_{t=1}^T\sum_{i}^N\sum_{j\notin\mathcal{N}_i} w_{t, j'}(\mathbf{x}_i'^\mathcal{H}, \mathbf{x}_j'^\mathcal{H}).
\end{aligned}
\end{equation}
Next, we center the above loss to each node. For each node, the constraint comes from two aspects: (1) temporal homophily loss, which minimizes the hyperbolic distance between the node and its positive neighbors in all timestamps, and maximizes the distance between the node and the sampled negative neighbors, where the weights are determined by $\frac{1}{|\mathcal{E}_t|}$; (2) consistency constraint between the same node over two consecutive timestamps, that is:
\begin{equation}
\begin{aligned}
\mathcal{L}&
    =\sum_{i}^N\sum_{t=1}^T\left(\sum_{j\in\mathcal{N}(i)} \frac{1}{|\mathcal{E}_t|} d^\mathcal{H}(\mathbf{x}_i^\mathcal{H}, \mathbf{x}_j^\mathcal{H})-\sum_{j\notin\mathcal{N}(i)} \frac{1}{|\mathcal{E}_t|}(\mathbf{x}_{i}^\mathcal{H}, \mathbf{x}_{j'}^\mathcal{H})\right.\\
    &\quad\quad\quad\quad\quad\quad\quad\quad\quad\quad\quad\quad\quad\quad\quad\left.+\frac{\lambda}{|V|}d^\mathcal{H}(\mathbf{x}_{(t, i)}^\mathcal{H}, \mathbf{x}_{(t-1, i)}^\mathcal{H})\right).
\end{aligned}
\end{equation}

\end{proof}

\subsection{Experiment details}
\subsubsection{Data processing}
\label{appendix:data_processing}
Most of the data is in a timestamp format and we process it according to the physical meaning of the real world. The details are as follows:\\
\textbf{Enron}\footnote{\url{https://www.cs.cornell.edu/~arb/data/email-Enron/}} is constructed from emails exchanged by $184$ Ernon employees. The nodes represent the employees and the edges indicate the email interactions between them. We follow the same processing procedure as~\cite{hajiramezanali2019VGRNN} to obtain 10 snapshots. The network does not contain any node and edge information. \\
\textbf{COLAB}\footnote{\url{https://github.com/VGraphRNN/VGRNN/tree/master/data}} is an academic cooperation network, including the academic cooperation of $315$ researchers from 2000 to 2009. Each node on the graph represents an author, and an edge denotes a co-authorship relation. We split the dataset by year following~\cite{hajiramezanali2019VGRNN} and obtain 10 snapshots.\\
\textbf{FB}\footnote{\url{http://networkrepository.com/ia-facebook-wall-wosn-dir.php}} is a social network graph of Facebook Wall posts where each node is a user and each edge is the interaction related to their wall posts. We take the activates over the last three years in the dataset as $36$ snapshots. The FB dataset is associated with a large number of users but very sparse connections.\\
\textbf{HepPh}\footnote{\url{https://snap.stanford.edu/data/cit-HepPh.html}} is a citation network related to high energy physics phenomenology, which is collected from the e-print arXiv website. Each node represents a paper, and an edge represents one paper citing another. The data covers papers in the period between January 1993 to April 2003 (124 months in total). It is a directed graph network, but we learn and predict as if it was an undirected graph. According to the real physical meaning, we use three months of data per snapshot and use the last $36$ months as the full dataset in our work.\\
\textbf{AS733}\footnote{\url{https://snap.stanford.edu/data/as-733.html}} is an Internet router network, which is collected from the University of Oregon Route Views Project. This dataset contains $733$ instances and spans the time from November 8, 1997, to January 2, 2000, with an interval of $785$ days. We split the snapshots per day and select the last $30$ snapshots to use in this work. It is worth noting that this network is different from the citation networks where the nodes and edges only increase over time (i.e., no deleted edges or nodes), the AS733 data set also contains the removal of nodes and edges over time.\\
\textbf{DISEASE}\footnote{\url{https://github.com/HazyResearch/hgcn/tree/master/data/disease\_lp}} is a synthetic dataset based on the SIR disease spreading model~\cite{bjornstad2002dynamics}, which also feasible for the COVID-19 path, where each node represents a person, the node feature describes the symptom of a person, and the edges indicate the spreading relationship. We split the dataset by the time they appear, and there are a large number of unobserved nodes in the test set.

\subsubsection{Parameter settings}
Note that most of the benchmark datasets for dynamic graph embedding are only associated with topology. Enron and COLAB are associated with a small number of nodes, we use identity matrix as the node feature which is identical with the processing in~\cite{hajiramezanali2019VGRNN}. For the other datasets, i.e., HepPh, FB, and AS733, their node features are initialized by a dense vector with 128 dimensions using glorot's method~\cite{glorot2010understanding} and are set as trainable matrices. The DISEASE is associate with node feature and we directly use it in our work. We set the final embeddings dimension of all models as 16, and it needs to be clear that different embedding sizes can lead to different results, but our method can always achieve impressive results.

We set the number of GRU layers as 1 for all models if there is a recurrent unit (e.g., RNN, GRU, LSTM, HGRU) for a fair comparison. In HTGN, we set the number of historical windows in HTA as $4$ for DISEASE and $5$ for the other datasets. We did not do any heavy parameter tuning since our main work is to verify whether HTA is an effective using module in HTGN. The hyper-parameters of $r$ and $t$ in the Fermi-Dirac function are set as $2.0$ and $1.0$ which is a common choice as~\cite{hgcn2019}.

% \subsubsection{Additional experimental results on blablabla}
% \subsubsection{More experiential results}
% \begin{figure}[t]
% \centering
% \label{fig:ablations}
% \includegraphics[width=7cm]{figures/step_by_step_prediction/HepPh60_long_time_prediction.pdf}
% \caption{Comparison using HTC and not using HTC modules.}
% % \caption{Comparison using different embedding dimensions.}
% \end{figure}

%% file: main.bbl

\begin{thebibliography}{42}

%%% the Web address.

\ifx \showCODEN    \undefined \def \showCODEN     #1{\unskip}     \fi
\ifx \showDOI      \undefined \def \showDOI       #1{#1}\fi
\ifx \showISBNx    \undefined \def \showISBNx     #1{\unskip}     \fi
\ifx \showISBNxiii \undefined \def \showISBNxiii  #1{\unskip}     \fi
\ifx \showISSN     \undefined \def \showISSN      #1{\unskip}     \fi
\ifx \showLCCN     \undefined \def \showLCCN      #1{\unskip}     \fi
\ifx \shownote     \undefined \def \shownote      #1{#1}          \fi
\ifx \showarticletitle \undefined \def \showarticletitle #1{#1}   \fi
\ifx \showURL      \undefined \def \showURL       {\relax}        \fi
% invisible to TeX
\providecommand\bibfield[2]{#2}
\providecommand\bibinfo[2]{#2}
\providecommand\natexlab[1]{#1}
\providecommand\showeprint[2][]{arXiv:#2}

\bibitem[\protect\citeauthoryear{Aggarwal and Subbian}{Aggarwal and
  Subbian}{2014}]%
        {TG_survey:Evolutionary}
\bibfield{author}{\bibinfo{person}{Charu Aggarwal} {and}
  \bibinfo{person}{Karthik Subbian}.} \bibinfo{year}{2014}\natexlab{}.
\newblock \showarticletitle{Evolutionary network analysis: A survey}.
\newblock \bibinfo{journal}{\emph{ACM Computing Surveys (CSUR)}}
  \bibinfo{volume}{47}, \bibinfo{number}{1} (\bibinfo{year}{2014}),
  \bibinfo{pages}{1--36}.
\newblock


\bibitem[\protect\citeauthoryear{Bac{\'a}k}{Bac{\'a}k}{2014}]%
        {bacak2014computing}
\bibfield{author}{\bibinfo{person}{Miroslav Bac{\'a}k}.}
  \bibinfo{year}{2014}\natexlab{}.
\newblock \showarticletitle{Computing medians and means in Hadamard spaces}.
\newblock \bibinfo{journal}{\emph{SIAM Journal on Optimization}}
  \bibinfo{volume}{24}, \bibinfo{number}{3} (\bibinfo{year}{2014}),
  \bibinfo{pages}{1542--1566}.
\newblock


\bibitem[\protect\citeauthoryear{Bj{\o}rnstad, Finkenst{\"a}dt, and
  Grenfell}{Bj{\o}rnstad et~al\mbox{.}}{2002}]%
        {bjornstad2002dynamics}
\bibfield{author}{\bibinfo{person}{Ottar~N Bj{\o}rnstad},
  \bibinfo{person}{B{\"a}rbel~F Finkenst{\"a}dt}, {and}
  \bibinfo{person}{Bryan~T Grenfell}.} \bibinfo{year}{2002}\natexlab{}.
\newblock \showarticletitle{Dynamics of measles epidemics: estimating scaling
  of transmission rates using a time series SIR model}.
\newblock \bibinfo{journal}{\emph{Ecological monographs}} \bibinfo{volume}{72},
  \bibinfo{number}{2} (\bibinfo{year}{2002}), \bibinfo{pages}{169--184}.
\newblock


\bibitem[\protect\citeauthoryear{Bridson and Haefliger}{Bridson and
  Haefliger}{2013}]%
        {bridson2013metric}
\bibfield{author}{\bibinfo{person}{Martin~R Bridson} {and}
  \bibinfo{person}{Andr{\'e} Haefliger}.} \bibinfo{year}{2013}\natexlab{}.
\newblock \bibinfo{booktitle}{\emph{Metric spaces of non-positive curvature}}.
  Vol.~\bibinfo{volume}{319}.
\newblock \bibinfo{publisher}{Springer Science \& Business Media}.
\newblock


\bibitem[\protect\citeauthoryear{Bronstein, Bruna, LeCun, Szlam, and
  Vandergheynst}{Bronstein et~al\mbox{.}}{2017}]%
        {bronstein2017geometric}
\bibfield{author}{\bibinfo{person}{Michael~M Bronstein}, \bibinfo{person}{Joan
  Bruna}, \bibinfo{person}{Yann LeCun}, \bibinfo{person}{Arthur Szlam}, {and}
  \bibinfo{person}{Pierre Vandergheynst}.} \bibinfo{year}{2017}\natexlab{}.
\newblock \showarticletitle{Geometric deep learning: going beyond euclidean
  data}.
\newblock \bibinfo{journal}{\emph{IEEE Signal Processing Magazine}}
  \bibinfo{volume}{34}, \bibinfo{number}{4} (\bibinfo{year}{2017}),
  \bibinfo{pages}{18--42}.
\newblock


\bibitem[\protect\citeauthoryear{Chami, Ying, R{\'e}, and Leskovec}{Chami
  et~al\mbox{.}}{2019}]%
        {hgcn2019}
\bibfield{author}{\bibinfo{person}{Ines Chami}, \bibinfo{person}{Zhitao Ying},
  \bibinfo{person}{Christopher R{\'e}}, {and} \bibinfo{person}{Jure Leskovec}.}
  \bibinfo{year}{2019}\natexlab{}.
\newblock \showarticletitle{Hyperbolic graph convolutional neural networks}. In
  \bibinfo{booktitle}{\emph{NeurIPS}}. \bibinfo{pages}{4868--4879}.
\newblock


\bibitem[\protect\citeauthoryear{Chang, Liu, Wen, Li, Fang, Song, and Qi}{Chang
  et~al\mbox{.}}{2020}]%
        {tne:Neural_process}
\bibfield{author}{\bibinfo{person}{Xiaofu Chang}, \bibinfo{person}{Xuqin Liu},
  \bibinfo{person}{Jianfeng Wen}, \bibinfo{person}{Shuang Li},
  \bibinfo{person}{Yanming Fang}, \bibinfo{person}{Le Song}, {and}
  \bibinfo{person}{Yuan Qi}.} \bibinfo{year}{2020}\natexlab{}.
\newblock \showarticletitle{Continuous-Time Dynamic Graph Learning via Neural
  Interaction Processes}. In \bibinfo{booktitle}{\emph{CIKM}}.
  \bibinfo{pages}{145--154}.
\newblock


\bibitem[\protect\citeauthoryear{Cho, Van~Merri{\"e}nboer, Gulcehre, Bahdanau,
  Bougares, Schwenk, and Bengio}{Cho et~al\mbox{.}}{2014}]%
        {GRU}
\bibfield{author}{\bibinfo{person}{Kyunghyun Cho}, \bibinfo{person}{Bart
  Van~Merri{\"e}nboer}, \bibinfo{person}{Caglar Gulcehre},
  \bibinfo{person}{Dzmitry Bahdanau}, \bibinfo{person}{Fethi Bougares},
  \bibinfo{person}{Holger Schwenk}, {and} \bibinfo{person}{Yoshua Bengio}.}
  \bibinfo{year}{2014}\natexlab{}.
\newblock \showarticletitle{Learning phrase representations using RNN
  encoder-decoder for statistical machine translation}.
\newblock \bibinfo{journal}{\emph{arXiv preprint arXiv:1406.1078}}
  (\bibinfo{year}{2014}).
\newblock


\bibitem[\protect\citeauthoryear{Cui, Wu, Huang, and Wang}{Cui
  et~al\mbox{.}}{2019}]%
        {HCA}
\bibfield{author}{\bibinfo{person}{Qiang Cui}, \bibinfo{person}{Shu Wu},
  \bibinfo{person}{Yan Huang}, {and} \bibinfo{person}{Liang Wang}.}
  \bibinfo{year}{2019}\natexlab{}.
\newblock \showarticletitle{A hierarchical contextual attention-based network
  for sequential recommendation}.
\newblock \bibinfo{journal}{\emph{Neurocomputing}}  \bibinfo{volume}{358}
  (\bibinfo{year}{2019}), \bibinfo{pages}{141--149}.
\newblock


\bibitem[\protect\citeauthoryear{Dwibedi, Aytar, Tompson, Sermanet, and
  Zisserman}{Dwibedi et~al\mbox{.}}{2019}]%
        {dwibedi2019temporal}
\bibfield{author}{\bibinfo{person}{Debidatta Dwibedi}, \bibinfo{person}{Yusuf
  Aytar}, \bibinfo{person}{Jonathan Tompson}, \bibinfo{person}{Pierre
  Sermanet}, {and} \bibinfo{person}{Andrew Zisserman}.}
  \bibinfo{year}{2019}\natexlab{}.
\newblock \showarticletitle{Temporal cycle-consistency learning}. In
  \bibinfo{booktitle}{\emph{CVPR}}. \bibinfo{pages}{1801--1810}.
\newblock


\bibitem[\protect\citeauthoryear{Fout, Byrd, Shariat, and Ben-Hur}{Fout
  et~al\mbox{.}}{2017}]%
        {fout2017protein}
\bibfield{author}{\bibinfo{person}{Alex Fout}, \bibinfo{person}{Jonathon Byrd},
  \bibinfo{person}{Basir Shariat}, {and} \bibinfo{person}{Asa Ben-Hur}.}
  \bibinfo{year}{2017}\natexlab{}.
\newblock \showarticletitle{Protein interface prediction using graph
  convolutional networks}. In \bibinfo{booktitle}{\emph{NeurIPS}}.
  \bibinfo{pages}{6530--6539}.
\newblock


\bibitem[\protect\citeauthoryear{Fr{\'e}chet}{Fr{\'e}chet}{1948}]%
        {frechet1948elements}
\bibfield{author}{\bibinfo{person}{Maurice Fr{\'e}chet}.}
  \bibinfo{year}{1948}\natexlab{}.
\newblock \showarticletitle{Les {\'e}l{\'e}ments al{\'e}atoires de nature
  quelconque dans un espace distanci{\'e}}. In
  \bibinfo{booktitle}{\emph{Annales de l'institut Henri Poincar{\'e}}},
  Vol.~\bibinfo{volume}{10}. \bibinfo{pages}{215--310}.
\newblock


\bibitem[\protect\citeauthoryear{Ganea, B{\'e}cigneul, and Hofmann}{Ganea
  et~al\mbox{.}}{2018}]%
        {HNN}
\bibfield{author}{\bibinfo{person}{Octavian Ganea}, \bibinfo{person}{Gary
  B{\'e}cigneul}, {and} \bibinfo{person}{Thomas Hofmann}.}
  \bibinfo{year}{2018}\natexlab{}.
\newblock \showarticletitle{Hyperbolic neural networks}. In
  \bibinfo{booktitle}{\emph{NeurIPS}}. \bibinfo{pages}{5345--5355}.
\newblock


\bibitem[\protect\citeauthoryear{Glorot and Bengio}{Glorot and Bengio}{2010}]%
        {glorot2010understanding}
\bibfield{author}{\bibinfo{person}{Xavier Glorot} {and} \bibinfo{person}{Yoshua
  Bengio}.} \bibinfo{year}{2010}\natexlab{}.
\newblock \showarticletitle{Understanding the difficulty of training deep
  feedforward neural networks}. In \bibinfo{booktitle}{\emph{Proceedings of the
  thirteenth international conference on artificial intelligence and
  statistics}}. JMLR Workshop and Conference Proceedings,
  \bibinfo{pages}{249--256}.
\newblock


\bibitem[\protect\citeauthoryear{Gu, Sala, Gunel, and R{\'e}}{Gu
  et~al\mbox{.}}{2019}]%
        {gu2019learning}
\bibfield{author}{\bibinfo{person}{Albert Gu}, \bibinfo{person}{Frederic Sala},
  \bibinfo{person}{Beliz Gunel}, {and} \bibinfo{person}{Christopher R{\'e}}.}
  \bibinfo{year}{2019}\natexlab{}.
\newblock \showarticletitle{Learning mixed-curvature representations in product
  spaces}. In \bibinfo{booktitle}{\emph{ICLR}}.
\newblock


\bibitem[\protect\citeauthoryear{Gulcehre, Denil, Malinowski, Razavi, Pascanu,
  Hermann, Battaglia, Bapst, Raposo, Santoro, et~al\mbox{.}}{Gulcehre
  et~al\mbox{.}}{2019}]%
        {gulcehre2019hyperbolicAT}
\bibfield{author}{\bibinfo{person}{Caglar Gulcehre}, \bibinfo{person}{Misha
  Denil}, \bibinfo{person}{Mateusz Malinowski}, \bibinfo{person}{Ali Razavi},
  \bibinfo{person}{Razvan Pascanu}, \bibinfo{person}{Karl~Moritz Hermann},
  \bibinfo{person}{Peter Battaglia}, \bibinfo{person}{Victor Bapst},
  \bibinfo{person}{David Raposo}, \bibinfo{person}{Adam Santoro},
  {et~al\mbox{.}}} \bibinfo{year}{2019}\natexlab{}.
\newblock \showarticletitle{Hyperbolic attention networks}. In
  \bibinfo{booktitle}{\emph{ICLR}}.
\newblock


\bibitem[\protect\citeauthoryear{Hajiramezanali, Hasanzadeh, Narayanan,
  Duffield, Zhou, and Qian}{Hajiramezanali et~al\mbox{.}}{2019}]%
        {hajiramezanali2019VGRNN}
\bibfield{author}{\bibinfo{person}{Ehsan Hajiramezanali},
  \bibinfo{person}{Arman Hasanzadeh}, \bibinfo{person}{Krishna Narayanan},
  \bibinfo{person}{Nick Duffield}, \bibinfo{person}{Mingyuan Zhou}, {and}
  \bibinfo{person}{Xiaoning Qian}.} \bibinfo{year}{2019}\natexlab{}.
\newblock \showarticletitle{Variational graph recurrent neural networks}. In
  \bibinfo{booktitle}{\emph{NeurIPS}}. \bibinfo{pages}{10701--10711}.
\newblock


\bibitem[\protect\citeauthoryear{Hochreiter and Schmidhuber}{Hochreiter and
  Schmidhuber}{1997}]%
        {LSTM}
\bibfield{author}{\bibinfo{person}{Sepp Hochreiter} {and}
  \bibinfo{person}{J{\"u}rgen Schmidhuber}.} \bibinfo{year}{1997}\natexlab{}.
\newblock \showarticletitle{Long short-term memory}.
\newblock \bibinfo{journal}{\emph{Neural computation}} \bibinfo{volume}{9},
  \bibinfo{number}{8} (\bibinfo{year}{1997}), \bibinfo{pages}{1735--1780}.
\newblock


\bibitem[\protect\citeauthoryear{Jonckheere, Lohsoonthorn, and
  Bonahon}{Jonckheere et~al\mbox{.}}{2008}]%
        {Gromov_hyperbolicity2}
\bibfield{author}{\bibinfo{person}{Edmond Jonckheere}, \bibinfo{person}{Poonsuk
  Lohsoonthorn}, {and} \bibinfo{person}{Francis Bonahon}.}
  \bibinfo{year}{2008}\natexlab{}.
\newblock \showarticletitle{Scaled Gromov hyperbolic graphs}.
\newblock \bibinfo{journal}{\emph{Journal of Graph Theory}}
  \bibinfo{volume}{57}, \bibinfo{number}{2} (\bibinfo{year}{2008}),
  \bibinfo{pages}{157--180}.
\newblock


\bibitem[\protect\citeauthoryear{Kipf and Welling}{Kipf and Welling}{2016}]%
        {kipf2016variational}
\bibfield{author}{\bibinfo{person}{Thomas~N Kipf} {and} \bibinfo{person}{Max
  Welling}.} \bibinfo{year}{2016}\natexlab{}.
\newblock \showarticletitle{Variational graph auto-encoders}.
\newblock \bibinfo{journal}{\emph{Bayesian Deep Learning Workshop (NIPS 2016)}}
  (\bibinfo{year}{2016}).
\newblock


\bibitem[\protect\citeauthoryear{Kipf and Welling}{Kipf and Welling}{2017}]%
        {gcn2017}
\bibfield{author}{\bibinfo{person}{Thomas~N Kipf} {and} \bibinfo{person}{Max
  Welling}.} \bibinfo{year}{2017}\natexlab{}.
\newblock \showarticletitle{Semi-Supervised Classification with Graph
  Convolutional Networks}. In \bibinfo{booktitle}{\emph{ICLR}}.
\newblock


\bibitem[\protect\citeauthoryear{Krioukov, Papadopoulos, Kitsak, Vahdat, and
  Bogun{\'a}}{Krioukov et~al\mbox{.}}{2010}]%
        {2010hyperbolic}
\bibfield{author}{\bibinfo{person}{Dmitri Krioukov},
  \bibinfo{person}{Fragkiskos Papadopoulos}, \bibinfo{person}{Maksim Kitsak},
  \bibinfo{person}{Amin Vahdat}, {and} \bibinfo{person}{Mari{\'a}n
  Bogun{\'a}}.} \bibinfo{year}{2010}\natexlab{}.
\newblock \showarticletitle{Hyperbolic geometry of complex networks}.
\newblock \bibinfo{journal}{\emph{Physical Review E}} \bibinfo{volume}{82},
  \bibinfo{number}{3} (\bibinfo{year}{2010}), \bibinfo{pages}{036106}.
\newblock


\bibitem[\protect\citeauthoryear{Liu, Wang, Hu, Duan, and Kot}{Liu
  et~al\mbox{.}}{2017}]%
        {liu2017global}
\bibfield{author}{\bibinfo{person}{Jun Liu}, \bibinfo{person}{Gang Wang},
  \bibinfo{person}{Ping Hu}, \bibinfo{person}{Ling-Yu Duan}, {and}
  \bibinfo{person}{Alex~C Kot}.} \bibinfo{year}{2017}\natexlab{}.
\newblock \showarticletitle{Global context-aware attention lstm networks for 3d
  action recognition}. In \bibinfo{booktitle}{\emph{CVPR}}.
  \bibinfo{pages}{1647--1656}.
\newblock


\bibitem[\protect\citeauthoryear{Liu, Nickel, and Kiela}{Liu
  et~al\mbox{.}}{2019a}]%
        {liu2019HGNN}
\bibfield{author}{\bibinfo{person}{Qi Liu}, \bibinfo{person}{Maximilian
  Nickel}, {and} \bibinfo{person}{Douwe Kiela}.}
  \bibinfo{year}{2019}\natexlab{a}.
\newblock \showarticletitle{Hyperbolic graph neural networks}. In
  \bibinfo{booktitle}{\emph{NeurIPS}}. \bibinfo{pages}{8230--8241}.
\newblock


\bibitem[\protect\citeauthoryear{Liu, Shi, Pierce, and Ren}{Liu
  et~al\mbox{.}}{2019b}]%
        {liu2019characterizing}
\bibfield{author}{\bibinfo{person}{Yozen Liu}, \bibinfo{person}{Xiaolin Shi},
  \bibinfo{person}{Lucas Pierce}, {and} \bibinfo{person}{Xiang Ren}.}
  \bibinfo{year}{2019}\natexlab{b}.
\newblock \showarticletitle{Characterizing and forecasting user engagement with
  in-app action graph: A case study of snapchat}. In
  \bibinfo{booktitle}{\emph{KDD}}. \bibinfo{pages}{2023--2031}.
\newblock


\bibitem[\protect\citeauthoryear{Lozier}{Lozier}{2003}]%
        {lozier2003nist}
\bibfield{author}{\bibinfo{person}{Daniel~W Lozier}.}
  \bibinfo{year}{2003}\natexlab{}.
\newblock \showarticletitle{NIST digital library of mathematical functions}.
\newblock \bibinfo{journal}{\emph{Annals of Mathematics and Artificial
  Intelligence}} \bibinfo{volume}{38}, \bibinfo{number}{1}
  (\bibinfo{year}{2003}), \bibinfo{pages}{105--119}.
\newblock


\bibitem[\protect\citeauthoryear{Narayan and Saniee}{Narayan and
  Saniee}{2011}]%
        {Gromov_hyperbolicity1}
\bibfield{author}{\bibinfo{person}{Onuttom Narayan} {and} \bibinfo{person}{Iraj
  Saniee}.} \bibinfo{year}{2011}\natexlab{}.
\newblock \showarticletitle{Large-scale curvature of networks}.
\newblock \bibinfo{journal}{\emph{Physical Review E}} \bibinfo{volume}{84},
  \bibinfo{number}{6} (\bibinfo{year}{2011}), \bibinfo{pages}{066108}.
\newblock


\bibitem[\protect\citeauthoryear{Nguyen, Lee, Rossi, Ahmed, Koh, and
  Kim}{Nguyen et~al\mbox{.}}{2018}]%
        {tne:ctdn}
\bibfield{author}{\bibinfo{person}{Giang~Hoang Nguyen},
  \bibinfo{person}{John~Boaz Lee}, \bibinfo{person}{Ryan~A Rossi},
  \bibinfo{person}{Nesreen~K Ahmed}, \bibinfo{person}{Eunyee Koh}, {and}
  \bibinfo{person}{Sungchul Kim}.} \bibinfo{year}{2018}\natexlab{}.
\newblock \showarticletitle{Continuous-time dynamic network embeddings}. In
  \bibinfo{booktitle}{\emph{WWW}}. \bibinfo{pages}{969--976}.
\newblock


\bibitem[\protect\citeauthoryear{Nickel and Kiela}{Nickel and Kiela}{2017}]%
        {nickel2017poincare}
\bibfield{author}{\bibinfo{person}{Maximillian Nickel} {and}
  \bibinfo{person}{Douwe Kiela}.} \bibinfo{year}{2017}\natexlab{}.
\newblock \showarticletitle{Poincar{\'e} embeddings for learning hierarchical
  representations}. In \bibinfo{booktitle}{\emph{NeurIPS}}.
  \bibinfo{pages}{6338--6347}.
\newblock


\bibitem[\protect\citeauthoryear{Nickel and Kiela}{Nickel and Kiela}{2018}]%
        {nickel2018learning}
\bibfield{author}{\bibinfo{person}{Maximillian Nickel} {and}
  \bibinfo{person}{Douwe Kiela}.} \bibinfo{year}{2018}\natexlab{}.
\newblock \showarticletitle{Learning Continuous Hierarchies in the Lorentz
  Model of Hyperbolic Geometry}. In \bibinfo{booktitle}{\emph{ICML}}.
  \bibinfo{pages}{3779--3788}.
\newblock


\bibitem[\protect\citeauthoryear{Pareja, Domeniconi, Chen, Ma, Suzumura,
  Kanezashi, Kaler, Schardl, and Leiserson}{Pareja et~al\mbox{.}}{2020}]%
        {pareja2020evolvegcn}
\bibfield{author}{\bibinfo{person}{Aldo Pareja}, \bibinfo{person}{Giacomo
  Domeniconi}, \bibinfo{person}{Jie Chen}, \bibinfo{person}{Tengfei Ma},
  \bibinfo{person}{Toyotaro Suzumura}, \bibinfo{person}{Hiroki Kanezashi},
  \bibinfo{person}{Tim Kaler}, \bibinfo{person}{Tao~B Schardl}, {and}
  \bibinfo{person}{Charles~E Leiserson}.} \bibinfo{year}{2020}\natexlab{}.
\newblock \showarticletitle{EvolveGCN: Evolving Graph Convolutional Networks
  for Dynamic Graphs}. In \bibinfo{booktitle}{\emph{AAAI}}.
  \bibinfo{pages}{5363--5370}.
\newblock


\bibitem[\protect\citeauthoryear{Sala, De~Sa, Gu, and Re}{Sala
  et~al\mbox{.}}{2018}]%
        {sala2018representation}
\bibfield{author}{\bibinfo{person}{Frederic Sala}, \bibinfo{person}{Chris
  De~Sa}, \bibinfo{person}{Albert Gu}, {and} \bibinfo{person}{Christopher Re}.}
  \bibinfo{year}{2018}\natexlab{}.
\newblock \showarticletitle{Representation Tradeoffs for Hyperbolic
  Embeddings}. In \bibinfo{booktitle}{\emph{ICML}}.
  \bibinfo{pages}{4460--4469}.
\newblock


\bibitem[\protect\citeauthoryear{Sankar, Wu, Gou, Zhang, and Yang}{Sankar
  et~al\mbox{.}}{2020}]%
        {sankar2020dysat}
\bibfield{author}{\bibinfo{person}{Aravind Sankar}, \bibinfo{person}{Yanhong
  Wu}, \bibinfo{person}{Liang Gou}, \bibinfo{person}{Wei Zhang}, {and}
  \bibinfo{person}{Hao Yang}.} \bibinfo{year}{2020}\natexlab{}.
\newblock \showarticletitle{DySAT: Deep Neural Representation Learning on
  Dynamic Graphs via Self-Attention Networks}. In
  \bibinfo{booktitle}{\emph{WSDM}}. \bibinfo{pages}{519--527}.
\newblock


\bibitem[\protect\citeauthoryear{Seo, Defferrard, Vandergheynst, and
  Bresson}{Seo et~al\mbox{.}}{2018}]%
        {tgn:GRUGCN}
\bibfield{author}{\bibinfo{person}{Youngjoo Seo}, \bibinfo{person}{Micha{\"e}l
  Defferrard}, \bibinfo{person}{Pierre Vandergheynst}, {and}
  \bibinfo{person}{Xavier Bresson}.} \bibinfo{year}{2018}\natexlab{}.
\newblock \showarticletitle{Structured sequence modeling with graph
  convolutional recurrent networks}. In \bibinfo{booktitle}{\emph{ICONIP}}.
  Springer, \bibinfo{pages}{362--373}.
\newblock


\bibitem[\protect\citeauthoryear{Skarding, Gabrys, and Musial}{Skarding
  et~al\mbox{.}}{2020}]%
        {TG_survey:foundations}
\bibfield{author}{\bibinfo{person}{Joakim Skarding}, \bibinfo{person}{Bogdan
  Gabrys}, {and} \bibinfo{person}{Katarzyna Musial}.}
  \bibinfo{year}{2020}\natexlab{}.
\newblock \showarticletitle{Foundations and modelling of dynamic networks using
  Dynamic Graph Neural Networks: A survey}.
\newblock \bibinfo{journal}{\emph{arXiv preprint arXiv:2005.07496}}
  (\bibinfo{year}{2020}).
\newblock


\bibitem[\protect\citeauthoryear{Trivedi, Farajtabar, Biswal, and Zha}{Trivedi
  et~al\mbox{.}}{2019}]%
        {tne:dyrep}
\bibfield{author}{\bibinfo{person}{Rakshit Trivedi}, \bibinfo{person}{Mehrdad
  Farajtabar}, \bibinfo{person}{Prasenjeet Biswal}, {and}
  \bibinfo{person}{Hongyuan Zha}.} \bibinfo{year}{2019}\natexlab{}.
\newblock \showarticletitle{Dyrep: Learning representations over dynamic
  graphs}. In \bibinfo{booktitle}{\emph{ICLR}}.
\newblock


\bibitem[\protect\citeauthoryear{Wang, Jabri, and Efros}{Wang
  et~al\mbox{.}}{2019}]%
        {wang2019learning}
\bibfield{author}{\bibinfo{person}{Xiaolong Wang}, \bibinfo{person}{Allan
  Jabri}, {and} \bibinfo{person}{Alexei~A Efros}.}
  \bibinfo{year}{2019}\natexlab{}.
\newblock \showarticletitle{Learning correspondence from the cycle-consistency
  of time}. In \bibinfo{booktitle}{\emph{CVPR}}. \bibinfo{pages}{2566--2576}.
\newblock


\bibitem[\protect\citeauthoryear{Yang, Meng, and King}{Yang
  et~al\mbox{.}}{2020}]%
        {FeatureNorm2020}
\bibfield{author}{\bibinfo{person}{Menglin Yang}, \bibinfo{person}{Ziqiao
  Meng}, {and} \bibinfo{person}{Irwin King}.} \bibinfo{year}{2020}\natexlab{}.
\newblock \showarticletitle{FeatureNorm: L2 Feature Normalization for Dynamic
  Graph Embedding}. In \bibinfo{booktitle}{\emph{2020 IEEE International
  Conference on Data Mining (ICDM)}}. \bibinfo{pages}{731--740}.
\newblock
\urldef\tempurl%
\url{https://doi.org/10.1109/ICDM50108.2020.00082}
\showDOI{\tempurl}


\bibitem[\protect\citeauthoryear{Ying, You, Morris, Ren, Hamilton, and
  Leskovec}{Ying et~al\mbox{.}}{2018}]%
        {ying2018hierarchical}
\bibfield{author}{\bibinfo{person}{Zhitao Ying}, \bibinfo{person}{Jiaxuan You},
  \bibinfo{person}{Christopher Morris}, \bibinfo{person}{Xiang Ren},
  \bibinfo{person}{Will Hamilton}, {and} \bibinfo{person}{Jure Leskovec}.}
  \bibinfo{year}{2018}\natexlab{}.
\newblock \showarticletitle{Hierarchical graph representation learning with
  differentiable pooling}. In \bibinfo{booktitle}{\emph{NeurIPS}}.
  \bibinfo{pages}{4800--4810}.
\newblock


\bibitem[\protect\citeauthoryear{Zhang, Wang, Jiang, Shi, and Ye}{Zhang
  et~al\mbox{.}}{2019}]%
        {zhang2019hyperbolic}
\bibfield{author}{\bibinfo{person}{Yiding Zhang}, \bibinfo{person}{Xiao Wang},
  \bibinfo{person}{Xunqiang Jiang}, \bibinfo{person}{Chuan Shi}, {and}
  \bibinfo{person}{Yanfang Ye}.} \bibinfo{year}{2019}\natexlab{}.
\newblock \showarticletitle{Hyperbolic graph attention network}. In
  \bibinfo{booktitle}{\emph{AAAI}}.
\newblock


\bibitem[\protect\citeauthoryear{Zhao, Song, Zhang, Liu, Wang, Lin, Deng, and
  Li}{Zhao et~al\mbox{.}}{2019}]%
        {zhao2019tgcn}
\bibfield{author}{\bibinfo{person}{Ling Zhao}, \bibinfo{person}{Yujiao Song},
  \bibinfo{person}{Chao Zhang}, \bibinfo{person}{Yu Liu}, \bibinfo{person}{Pu
  Wang}, \bibinfo{person}{Tao Lin}, \bibinfo{person}{Min Deng}, {and}
  \bibinfo{person}{Haifeng Li}.} \bibinfo{year}{2019}\natexlab{}.
\newblock \showarticletitle{T-gcn: A temporal graph convolutional network for
  traffic prediction}.
\newblock \bibinfo{journal}{\emph{IEEE Transactions on Intelligent
  Transportation Systems}} (\bibinfo{year}{2019}).
\newblock


\bibitem[\protect\citeauthoryear{Zhu, Pan, Zhou, Wu, Cao, and Wang}{Zhu
  et~al\mbox{.}}{2020}]%
        {zhu2020gil}
\bibfield{author}{\bibinfo{person}{Shichao Zhu}, \bibinfo{person}{Shirui Pan},
  \bibinfo{person}{Chuan Zhou}, \bibinfo{person}{Jia Wu},
  \bibinfo{person}{Yanan Cao}, {and} \bibinfo{person}{Bin Wang}.}
  \bibinfo{year}{2020}\natexlab{}.
\newblock \showarticletitle{Graph Geometry Interaction Learning}. In
  \bibinfo{booktitle}{\emph{NeurIPS}}, Vol.~\bibinfo{volume}{33}.
\newblock


\end{thebibliography}
